\documentclass[12pt,preprint]{aastex}
\usepackage{graphicx}
\shorttitle{Eccentricity of extrasolar planets}
\shortauthors{F. Namouni}

\begin{document}
\title{On the origin of the eccentricities of extrasolar planets}
\author{Fathi Namouni}
\affil{CNRS,  Observatoire de la C\^ote d'Azur, BP 4229, 06304 Nice, France}
\email{namouni@obs-nice.fr}

\begin{abstract}
We develop a phenomenological theory that aims to account for the origin of
the large eccentricities of extrasolar planets and that of the small
eccentricities in the solar system, the preference for apsidal alignment in
non-resonant multiplanet systems, and the origin of the similarities in the
eccentricity distribution of extra-solar planets and that of spectroscopic
binary stars. We show that if a physical process is weakly dependent on the
local dynamics of the companion and imparts a small relative acceleration to
the star-companion system, the eccentricity of the companion's orbit is
excited to large values depending on the direction and duration of
acceleration.  A natural candidate for such processes are asymmetric stellar
jets and star-disk winds.  When the acceleration originates from a precessing
jet, large eccentricities can be excited by the resonance of the jet's
precession frequency with the induced acceleration's excitation frequency even
for nearly perpendicular jets.  Precession also reduces the eccentricity
amplitude far inside the resonance radius. The acceleration's strength is best
constrained in multiplanet systems because of the companions' mutual
gravitational perturbations, while the acceleration's duration is bounded by
the condition that the residual velocity imparted to the star is smaller than
the stellar velocity dispersion in the Galaxy. In the outer parts of the
star-companion system where the acceleration excitation time is comparable to
or smaller than the orbital period, significant radial migration takes place
which may have important consequences for the dynamics of the minor body
populations in the solar system. The theory is illustrated with the
$\upsilon$~Andromedae binary system.
\end{abstract}

\section{ Introduction}
Since the discovery of the first Jupiter-like planet around 51 Pegasi in 1995,
the number of extrasolar planets has exceeded 150, some of which reside in 14
multiplanet systems \citep{b20}. Extrasolar planetary orbits differ from the
solar system's in two ways: first is their semi-major axis
distribution. Planets can orbit their parent stars much closer than the tenth
of Mercury's distance to the Sun.  Second, at a median value of $e=0.27$,
their orbital eccentricities are so high that the only solar system analogs
are small bodies such as asteroids and Kuiper belt objects whose orbits have
been strongly stirred up by the gas giants. These differences are the more
puzzling as it has been recognized that the distribution of semi-major axes
and eccentricities of extrasolar planets resemble those of spectroscopic
binary star systems, a finding that may hint to common formation or excitation
mechanisms \citep{b22}.

Current theories of planetary formation suggest that Jupiter-like planets form
in a disk of gas and dust orbiting a parent star.  Producing planets with
small semi-major axes is a natural outcome of the transfer of angular momentum
between the disk and the planet \citep{b25}. The overall effect of this
transfer causes the planet's orbit to shrink around the star.  Radial
migration can be stopped by the tidal interaction between the planet and the
star \citep{b12} or through photo-evaporation of the disk's portion exterior
to the planet's orbit and that is responsible for its inward motion
\citep{b6}.

The presence of large eccentricities is perhaps the most surprising feature of
extrasolar systems as its unexplained origin contrasts with the known ability
of the gas disk to force migration.  The transfer of angular momentum between
the disk and the planet at the locations of their orbital resonances damps the
planetary eccentricity so that the natural outcome of this interaction is a
planet on a circular orbit \citep{b29}. This lead to the investigation of
other possibilities: secular perturbations by distant companions
\citep{b8,b23}, resonant interactions within a multiplanet system
\citep{b3,b11}, and encounters of a spatially extended multiplanet system
with passing stars \citep{b27}.  The advantages and drawbacks of these models
have been reviewed by \citet{b24}.  The main recurring observation is that
these models provide a solution for some specific planetary system. Moreover,
they do not explain the apparent uniqueness of the small eccentricities of the
solar system's planets, the occurrence of the largest eccentricities in some
multiplanet systems for the outer more massive companions, and the origin of
the similarity with spectroscopic binaries. In this regard, it is conceivable
to assume that planets formed from a collapsing proto-stellar cloud but this
would solve the problem only for the more massive known planets \citep{b19}.

In this work, we aim to explain these observations within a single framework.
We choose the premise of standard planetary formation theory that the natural
outcome of planetesimal accumulation and gas accretion are planets on
co-planar circular orbits. To account for the similarity of the eccentricity
distributions of extrasolar planets and spectroscopic binaries, we consider
the excitation caused by physical processes that are weakly dependent on the
local dynamics of the companion and which can therefore be modeled by an
acceleration that does not depend explicitly on the relative position and
velocity of the companion. In section 2, we study the general properties of
such an acceleration and show that stellar mass loss phenomena such as jets
and winds are natural candidates for the excitation process. However, we do
not specialize in the case of stellar jets because we prefer to present this
theory as a phenomenological one. The reason is twofold: first, the current
knowledge of stellar jets and winds both theoretical and observational is not
sufficient to issue a definitive verdict on their role in the eccentricity
excitation. Second, the dynamical mechanisms that arise because of such
accelerations and that lead to eccentricity excitation and apsidal alignment
are not concerned directly with the details of jet and wind generation. In
section 3, we demonstrate the basic eccentricity excitation mechanism for the
simplest finite duration acceleration, one whose direction is constant with
respect to the star-companion system.  In the remainder of the paper, we
consider the various constraints on the basic acceleration model of section 3
set by: the acceleration's precession (section 4), the radial migration of the
companion (section 5), the companion's orbital stability and the consistency
with the host star's galactic motion (section 6), the secular perturbations
from mutual gravitational interactions in multiplanet systems (section 7), and
the presence of distant companions (section 8).  The application of this
theory to the solar system is further discussed in section 8. Section 9
contains concluding remarks.

\section{Phenomenology}
In this section, we discuss the form of the eccentricity generating processes
that would apply equally to planetary and stellar companions. To this end, we
seek processes that are weakly dependent on the local orbital dynamics of the
companions. By this we mean that the interaction does not depend explicitly on
the relative position and relative velocity of the star-companion system.
With this in mind, we note that the simplest way to excite the eccentricity of
a companion's orbit is to subject it to a constant acceleration. Stable
excitation of the eccentricity is possible if the constant acceleration is
smaller than or comparable to the gravitational accelerations between the star
and its companions.  Writing the perturbing acceleration as ${\bf A}= A {\bf
u}$ where $A$ represents its magnitude and ${\bf u}$ is a unit vector, the
eccentricity excitation time is proportional to $v/A$, where $v$ is the
keplerian velocity of the companion. This determines two regions around the
host star whether the ratio of the excitation time to the orbital period is
larger than or comparable to unity. The former is located closer to the star
as the excitation time decreases with the relative distance of the star to the
companion. We call this region the secular excitation region as the
eccentricity increase builds up slowly after each rotation of the companion
around the star. The eccentricity excitation in this case is derived in
sections 3 and 4. The outer boundary of the secular excitation region is the
sudden excitation region where the excitation time is comparable to or smaller
than the orbital period. The dynamics in the sudden excitation region is
considered in section 6. With the acceleration magnitudes that we determine
below, the observed planetary companions fall into the secular region. Here
and if the perturbing acceleration were strictly constant, the corresponding
force would be conservative and the eccentricity excitation would simply
result in periodic oscillations between the original value, zero, and some
maximum value determined by the parameters of each individual system. We would
require that $A$ be finite for a duration comparable to or smaller than the
oscillation period induced by the acceleration in order to ensure a finite
residual eccentricity.  In the secular region, the duration is naturally
larger than the orbital period of the planet. In the sudden excitation region,
the duration of excitation can be comparable to or smaller than the orbital
period of distant companions and the outer minor bodies of the planetary
system in order to ensure stability (see sections 6 and 8).  We can obtain a
minimal estimate of the acceleration's maximal amplitude, $A_0$, in the
secular region by requiring that the eccentricity excitation time, be shorter
than the age of the planetary system $ \sim 10^9$ years. This yields: $A_0>
3\times 10^{-16}(v/10\,\mbox{km\,s}^{-1})\ \mbox{km\,s}^{-2}$.

The direction of acceleration, ${\bf u}$, needs to be specified in an inertial
frame as the acceleration does not excite the eccentricity of a companion
located in the secular region if it is perpendicular to its orbital plane.
The knowledge of the current state of the solar system suggests two options: a
local one related to the mean orbital plane of the planetary system, and a
global one related to the motion of the system in the Galaxy. For the local
option, the vector ${\bf u}$ is referred to the total angular momentum of the
planetary system or the host star's rotation axis. In the solar system, we
would expect ${\bf u}$ to be closest to the angular momentum direction because
of the small eccentricities of Jupiter and Saturn. The direction of ${\bf u}$
can be made independent of the planetary planes if it is referred to the Sun's
rotation axis which is inclined by 6$^\circ$ to Jupiter's orbital plane
normal. On the scale of a planetary system, the simplest way to achieve an
acceleration that is weakly dependent on the dynamics of the companion is by
stellar and disk mass loss.  The global option can be motivated by the
galactic configuration of the solar system: the planets' mean orbital plane is
inclined to the Sun's orbital plane in the Galaxy by the large value of 60
degrees. Jupiter and Saturn have small eccentricities compared to most of the
known giant planets that are not located in their parent stars' tidal
zones. The direction of the acceleration can then be tied to that of the
star's motion in the Galaxy with the expectation that the largest
eccentricities correspond to the situation where the planetary orbital plane
is coincident with the parent star's galactic motion plane. The simplest two
options for ${\bf u}$ are a constant vector that lies in the galactic
mid-plane and does not depend on the star's motion, and a vector directly
related to the star's motion such as its velocity in the Galaxy, ${\bf v}_g$,
or the orthogonal thereof ${\bf b}\times {\bf v}_g$ where ${\bf b}$ is a
constant vector. On the larger scale of the solar neighborhood, there are no
obvious processes that do not depend explicitly on the relative position and
relative velocity of the star-companion system and that would excite the
companion's eccentricity.

Jets and star-disk winds are ubiquitous features of star formation and
star-disk interaction \citep{b52,b53} and are natural candidates for the
origin of the perturbing acceleration.  Inferred mass loss rates for known
young T Tauri stars lie in the range $\sim 10^{-8}M_\odot\, \mbox{\rm
year}^{-1}$ to $10^{-10}M_\odot\, \mbox{\rm year}^{-1}$ and may be two orders
of magnitude larger depending on the way the rate is measured from the
luminosity of forbidden lines \citep{b46,b48,b47}.  The mass loss process
needs to be asymmetric with respect to the companion's orbital plane in order
to produce a residual acceleration. Interestingly, a number of bipolar jets
from young stars \citep{b50,b49} are known to be asymmetric as the velocities
of the jet and counter-jet differ by about a factor of 2. Mass loss processes
in young stars therefore yield maximal accelerations:
\begin{equation}
A_0 \sim  10^{-13}\, 
\left(\frac{\dot M}{10^{-8} M_\odot\, \mbox{\rm  year}^{-1}}\right)\, 
\left(\frac{v_e}{300 \,\mbox{\rm km\,s}^{-1}}\right)\,
\left(\frac{M_\odot}{M}\right) \mbox{\rm km\,s}^{-2}. 
\end{equation}
where $M$ is the stellar mass and $v_e$ is the outflow's high velocity
component. The implicit proportionality constant depends on the relative mass
loss ratios $\dot M$ and ejection velocities of the jet and counter-jet. 

Integrated over the duration of acceleration, the star acquires a
residual velocity that must be smaller than its galactic velocity else the
star is ejected. The residual velocity is further bound by the known random
component of the stellar motion. In section 8, we further discuss the mass
loss rate values after we consider the constraints obtained from the
companion's orbital stability and from the eccentricity excitation of
multiplanet systems.  Another requirement for stellar and disk mass loss to be
efficient in the secular region is that the momentum communicated to the star
be inclined with respect to the companion's orbital plane as it is the
component of the acceleration that lies in the companion's orbital plane that
excites the eccentricity. Studies of molecular outflows suggest that some jets
precess over timescales from $10^2$ to $10^4$ years \citep{b55,b54,b51}.  As
most T Tauri stars are known to be in multiple systems, a possible way to
achieve a jet precession is by warping the accreting disk's plane through the
gravitational perturbation of a stellar companion on an inclined orbit
\citep{b51}.  This would not imply that jet acceleration may only explain the
large eccentricities of planetary companions in binary systems.  We show in
sections 6 and 8 that the acceleration may force the outward migration and
possibly the ejection of the stellar companion.

In this paper, we do not specialize in a specific acceleration generating
process and therefore we do not enter into such specific details as asymmetric
jet generation and precession by an accreting circumstellar disk tidally
interacting with a stellar binary component.  We study the eccentricity
excitation by representing the perturbing process by a finite duration
acceleration that is independent of the relative position and velocity of the
system. In this respect, the application of this theory to the efficiency of
stellar jets in exciting eccentricities is phenomenological and yields useful
bounds on the star-disk mass loss rate. We point out that the excitation
process that is modeled by this acceleration may depend implicitly on the
relative position and velocity of the companion. For instance, the accretion
that powers the stellar jet is driven by viscosity which can itself be driven
by the planetary embryos through the shocking of the pressure waves they
launch in the gas disk \citep{b57}. In this case, the acceleration from the
jet will depend implicitly on the companions' local dynamics.

The equation of motion is that of the two-body problem modified as follows:
\begin{equation}
\frac{{\rm d} {\bf v}}{{\rm d} t}=-\frac{G(m+M)}{|{\bf x}|^3}\,{\bf
  x} +{\bf A}(t) \label{motion}
\end{equation}
where ${\bf A}$ is the relative acceleration acting on the star-companion
system, ${\bf x}$ and ${\bf v}$ are the relative position and velocity, $M$
and $m$ are the masses of the star and the companion. The acceleration is
finite over a typical duration that we call $\tau$. For simplicity, we choose
the amplitude of acceleration to be one of $A_1(t)=A_0H(t)\exp-t/\tau$ and
$A_2(t)=A_0/\cosh(t-t_0)/\tau$ where $H(t)$ is the Heaviside unit step
function and $A_0$ is the maximum acceleration amplitude.  We will use
the notation $A_0$ when we refer to a constant amplitude acceleration.  The
connection between the perturbing acceleration and the star's random motion in
the Galaxy makes the residual velocity, ${\bf V}$, a relevant parameter in the
problem; it is defined as:\begin{equation} {\bf
V}=\int_{-\infty}^{+\infty}{\bf A}(t)\, {\rm d}t.
\label{bigv}
\end{equation}
For a constant ${\bf u}$ and the two specific forms we use, $V_1=A_0\tau$ and
$V_2=\pi A_0\tau$.  We note that it is likely that the companion is subject to
radial migration due for instance to the tidal interaction with the
circumstellar disk.  The effect of migration on the eccentricity excitation is
discussed in section 5 where we modify (\ref{motion}) to account for migration
and show that our conclusions about the eccentricity remain valid in the
presence of migration. By using equation (\ref{motion}), we also neglect the
tidal effect of the Galactic potential; this is justified by the typical sizes
of the planetary and stellar systems considered here.

The eccentricity excitation time, $v/A_0$, allows us to split the excitation
problem into two cases: the first is when the variation of the direction of
${\bf A}$ is small on the scale $v/A_0$ in which case the direction of ${\bf
A}$ can be held constant to a good approximation. The eccentricity excitation
is demonstrated in this case in section 3. The second case is when the
direction of ${\bf A}$ varies on a timescale comparable or shorter than
$v/A_0$. This is the general case of precessing jets which is examined in
section 4.

\section{Eccentricity excitation}
Under the effect of a constant-direction acceleration, ${\bf A}=A(t){\bf u}$,
the variation of specific angular momentum, ${\bf h}={\bf x}\times {\bf v}$,
is $\dot {\bf h}={\bf x}\times{\bf A}$. As ${\bf u}$ is constant, only the
projection of the angular momentum along the direction of acceleration, ${\bf
h}\cdot {\bf u}$, is conserved. The rate of change of the angular momentum in
the direction orthogonal to ${\bf h}$ and ${\bf u}$ is obtained from $({\bf u}
\times {\bf h})\cdot \dot{\bf h}=A\ ({\bf u}\cdot{\bf x})\ ({\bf h}\cdot{\bf
u})$ and shows that if ${\bf u}$ lies in the orbital plane, the orbit does not
gain inclination with respect to its initial state.  To find the eccentricity
and inclination excitation of the companion's orbit in the secular excitation
region, we average the equations of motion and in particular the perturbing
acceleration over the fast orbital motion of the companion. This is valid
because the excitation time and the acceleration's duration are larger than
the orbital period of the companion. We can therefore assume that the
acceleration is independent of time in deriving the secular equations.  A
constant acceleration derives from the potential $R={\bf A}\cdot {\bf x}$. Its
average with respect to the orbital motion is derived in the Appendix and
reads:
\begin{equation}
\left<R\right> =-\frac{3}{2}
 a \, {\bf A}(t)\cdot {\bf e}=-
\frac{3}{2}
A(t) \, a e \sin(\varpi-\Omega) \sin I
\label{secpot}
\end{equation}
where ${\bf e}={\bf v}\times {\bf h}/G(m+M)-{\bf x}/|{\bf x}|$ is the
eccentricity vector of magnitude $e$ and $a$ is the orbital semi-major axis.
The last expression is obtained by choosing the $z$--axis along ${\bf u}$; in
this case, $\varpi$, $\Omega$ and $I$ are the longitude of pericenter, the
longitude of ascending node and the inclination of ${\bf h}$ with respect to
${\bf u}$.  The freedom of choosing a vector base in this plane results from
the conservation of ${\bf h}\cdot {\bf u}$ and leads to the appearance of the
combination $\omega=\varpi-\Omega$, the argument of pericenter.

\begin{figure}
\begin{center}\epsscale{.7}
\plotone{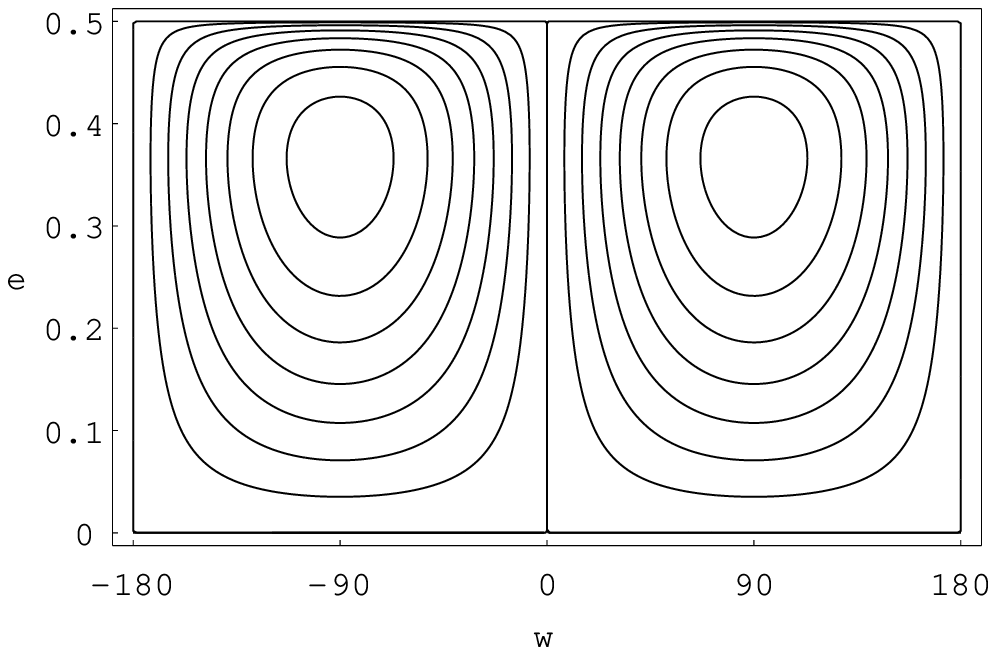}
\end{center}
\caption{Contour plots of the acceleration potential (\ref{secpot2}) in the
         eccentricity $e$ and argument of pericenter $\omega(^\circ)$
         plane. The direction of acceleration makes an angle $I_0=30^\circ$
         with respect to the companion's angular momentum vector. The time
         evolution of the two orbits ($e=0$, $\omega=0$) and ($e=0.3$,
         $\omega=90^\circ$) is shown in Figure (\ref{fig-2}).}
\label{fig-1}
\end{figure}

The conservation of ${\bf u}\cdot {\bf h}$ can be written as $\sqrt{1-e^2}
\cos I = \cos I_0$, where $I_0$ is the initial inclination of the circular
orbit, and allows us to reduce the problem to a single degree of freedom with
the potential:
\begin{equation}
\left< R\right>= -\frac{3}{2}
A\, a  \,
     \sqrt{\frac{\sin^2I_0-e^2}{1-e^2}}\, e \sin \omega. \label{secpot2}
\end{equation}
The eccentricity and pericenter evolution  is obtained from 
\begin{equation}
\dot e=-\frac{\sqrt{1-e^2}}{na^2\,e}\ \frac{\partial\left< R\right>}{\partial
     \omega}, \ \ \ \dot \omega=\frac{\sqrt{1-e^2}}{na^2\,e}\
     \frac{\partial\left< R\right>}{\partial e}, \label{edotomdot}
\end{equation} 
where $n=\sqrt{G(M+m)/a^3}$ is the companion's mean motion \citep{b56}.  In a
conservative system where $A(t)=A_0$ is constant, $e$ and $\omega$ follow
curves of constant $\left< R\right>$. There are equilibria at $\omega=\pm
90^\circ$ and $e=\sqrt{2} \sin (I_0/2)$ corresponding to
$I=\cos^{-1}(\sqrt{\cos I_0})$. The maximum value of $e$ is $\sin I_0$ and
corresponds to the cycle of initially circular orbits. In Figure
(\ref{fig-1}), we show the orbits of the conservative case for an inclination
$I_0=30^\circ.$

\begin{figure}\epsscale{.7}
\plotone{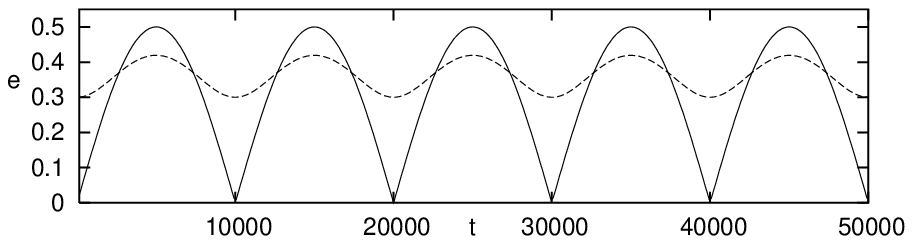}\\
\plotone{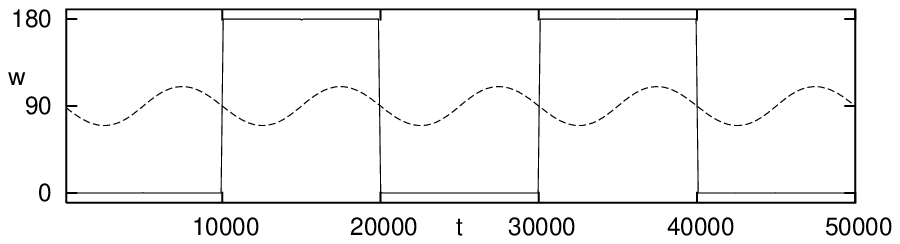}\\
\plotone{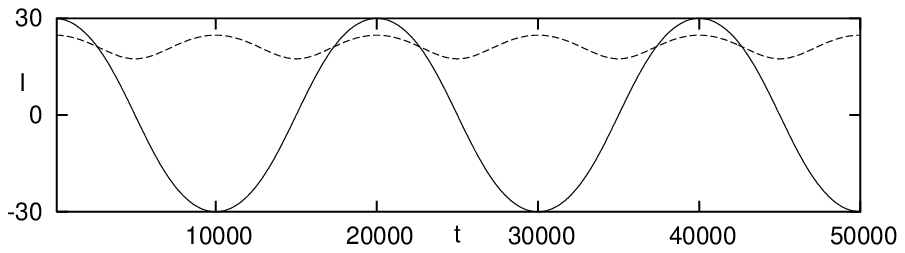}
\caption{Time evolution under a conservative acceleration. The eccentricity
         $e$, argument of pericenter $\omega(^\circ)$ and inclination
         $I(^\circ)$ are shown for an initially circular orbit $e=0$ (solid)
         and an orbit librating about the secular resonance $\omega=90^\circ$
         with an initial eccentricity $e=0.3$ (dashed).  The semi-major axis
         is identical for both orbits and is set to unity.  The acceleration
         corresponds to a period of $10^4$ years at 1\,AU. The plots were
         obtained by the numerical integration of the full equations of motion
         (\ref{motion}).}
\label{fig-2}
\end{figure}

\begin{figure}\epsscale{.7}
\plotone{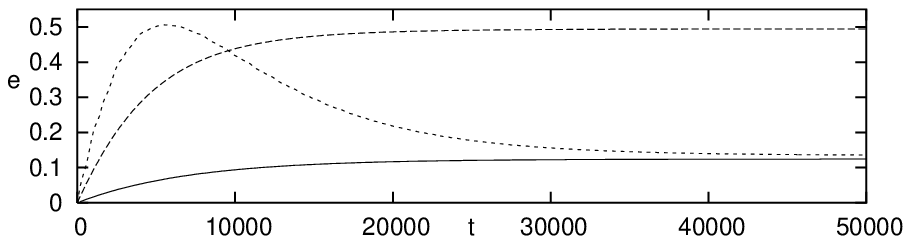}\\
\plotone{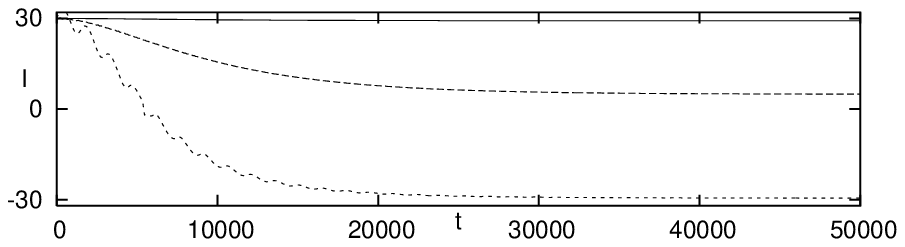}\\
\plotone{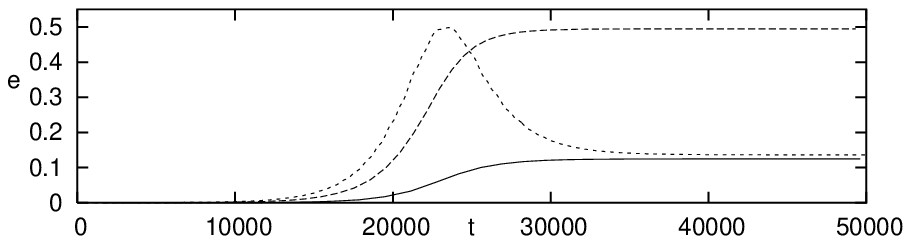}\\
\plotone{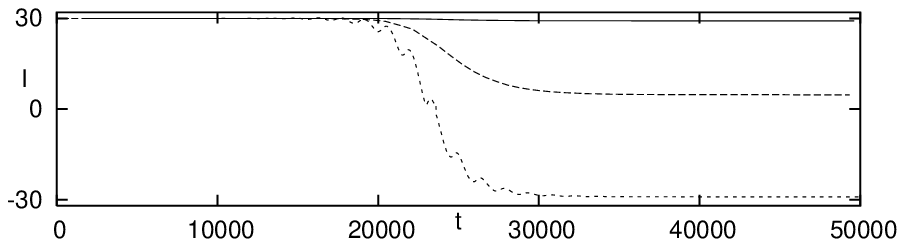}
\caption{Eccentricity excitation by time-dependent constant-direction
         accelerations. The full equations of motion (\ref{motion}) are
         integrated numerically with the forms $A_1$ (two upper panels) and
         $A_2$ (two lower panels) and $I_0=30^\circ$. The parameters are:
         $A_0=2.21\times 10^{-11}$\,km\,s$^{-2}$ and $t_0=2.3\times 10^{4}$
         years.  The oscillation period at 1\,AU is $1.11\times 10^5$ years.
         The timescale $\tau$ is chosen so that $V=$5\,km\,s$^{-1}$;
         $\tau=7200$ for $A_1$ and $\tau=2300$ years for $A_2$.  The curves
         correspond the semi-major axes: 1\,AU (solid), 32\,AU (dashed) and
         128\,AU (dotted). Note how the final $e$ and $I$ are equal under the
         two different accelerations at each semi-major axis.}
\label{fig-3}
\end{figure}

For initially circular orbits, the identity $\left< R\right>=0$ leads to
$\omega=0$ or $180^\circ$ throughout the evolution.  The eccentricity equation
then reads:
\begin{equation}
\dot e=\frac{3A(t) \epsilon}{2na}\ 
\,
     \sqrt{\sin^2I_0-e^2}, \label{edot}
\end{equation}
where $\epsilon$ is the sign of $\cos\omega$ which is set by the requirement
that $e\geq0$. For initially circular orbits, the discontinuous changes of
$\omega$ between 0 and 180$^\circ$ correspond to the crossing of the plane
$I=0$ during the oscillation of $I$ between $-I_0$ and $I_0$; it is the result
of the geometric requirement that $0\leq I\leq 180^\circ$ in standard
keplerian variables.  Finally, the eccentricity that is excited by the
perturbing acceleration is:
\begin{equation}
e(T)= \left|\sin\left[ \frac{3}{2na}\int_{-\infty}^{T}\,
 A(t)\,{\rm d}t\right]\ \sin I_0 \right|.
\label{eoft}
\end{equation}
The inclination is obtained from $\cos I= \cos I_0/\sqrt{1-e(T)^2}$.  In the
conservative case, $A(t)=A_0$, $e$ would oscillate between $0$ and $\sin I_0$
at the excitation frequency:
\begin{equation}
n_A=\frac{3\left|A_0\right|}{na}\,
  . \label{nsec}
\end{equation}
Examples of such oscillations that were obtained from the direct integration
of the full equations of motion are shown in Figure (\ref{fig-2}). The
agreement between the secular solution and the results numerical integration
is perfect and is due to the fact that $A$ is independent of the relative
position and velocity.  To optimize the excitation of a finite eccentricity
from an initially circular state, the duration of acceleration needs to be
smaller than half the oscillation period: $\tau<\pi na/3|A_0|$. A less
conservative criterion for eccentricity excitation is that the adiabatic
condition, $\tau\gg\pi na/3|A_0|$, that would ensure a long term decrease of
$e$ to zero is not satisfied. This allows the eccentricity to experience a few
oscillations before settling down to a finite value. For multiplanet systems,
this situation leads to strong interactions among the companions which can
lead to ejections offering a possible explanation for the existence of many
extrasolar planetary systems with one large planet.  Examples of eccentricity
excitation at three different semi-major axes (i.e. three different excitation
frequencies) are shown in Figure \ref{fig-3} for the two forms of the
acceleration's amplitude where $\tau$ has been chosen to yield the same
residual velocity $V$ and hence the same final eccentricities for each
semi-major axis. Figure \ref{fig-3} also illustrates the dependence of the
excitation amplitude on the ratio of the duration to the excitation time.  We
note that because eccentricity excitation in the secular region is a slow
process compared to the orbital time, the convolution of the dynamics under
the conservative acceleration $A_0$ with a finite time window has the effect
of shutting off the excitation at some eccentricity value depending on the
duration.

We conclude that the eccentricity $e$ can be excited up to $\sin I_0$ and is
largest if the initial orbital plane contains the direction of acceleration
($I_0=90^\circ$).  As $\left< R\right>=0$ for initially circular orbits, the
argument of pericenter, $\omega$, and the longitude of pericenter, $\Omega$,
remain at zero. We show in sections 7 and 8 that this forcing of the
pericentre to be aligned with the direction of acceleration favors apsidal
alignment in multiplanet system.  The inclination $I$ decreases from its
initial value pushing the orbital plane away from the direction of
acceleration.  We also note a interesting feature of this model that the final
eccentricity increases outward so that in multiplanet systems, the farthest
planets may have the largest eccentricities. This is a consequence of the
relative strength of the gravitational acceleration and the perturbing
acceleration as a function of distance.

\section{Precessing accelerations}
The inclination of the direction of acceleration ${\bf u}$ is crucial to the
excitation of eccentricity in the secular region. If the acceleration is due
to a stellar jet that precesses, we expect the excitation amplitude to depend
on the ratio of the excitation frequency $n_A$ to the precession frequency
$\Omega_A$ and that resonant forcing is possible when a match occurs, a
situation that is likely since $n_A$ is an increasing function of the
semi-major axis ${a}$. In the following, we solve the conservative excitation
problem with precession, that is we choose ${\bf A}= A_0\, {\bf u}(t)$ where
${\bf u}$ rotates at the rate $\Omega_A$.  For this problem, it is more
convenient to derive the eccentricity vector and angular momentum evolution
with secular perturbation theory in vector form \citep{b16} as:
\begin{eqnarray}
\frac{{\rm d}{\bf e}}{{\rm d}t}&=& \frac{1}{na^2}\,{\bf k}\times \nabla_{\bf e}
\left<R\right>  + \frac{1}{na^2}\,{\bf e}\times \nabla_{\bf k} 
\left<R\right> 
, \ \  \ \\
\frac{{\rm d}{\bf k}}{{\rm d}t}&=& \frac{1}{na^2}\,{\bf k}\times \nabla_{\bf k} 
\left<R\right>  + \frac{1}{na^2}\,{\bf e}\times \nabla_{\bf e} 
\left<R\right>, \label{milankovic}
\end{eqnarray}
where ${\bf k}={\bf h}/\sqrt{G(M+m)a}$ is the non-dimensional angular momentum
vector. Using the secular potential (\ref{secpot2}), these equations reduce
to:
\begin{eqnarray}
\frac{{\rm d}{\bf e}}{{\rm d}t}&=&- \frac{3A_0}{2na}\, {\bf k}\times {\bf u}, \ \
\frac{{\rm d}{\bf k}}{{\rm d}t}=- \frac{3A_0}{2na}\, {\bf e}\times {\bf u}
  \\
\frac{{\rm d}{\bf u}}{{\rm d}t}&=&  {\bf \Omega}_A\times {\bf u}, 
\label{milanexpand}
\end{eqnarray} 
where ${\bf \Omega}_A$ is the rotation vector associated with the precessing
acceleration. With precession, the constants of motion are given as:
\begin{equation}
-\frac{3A_0}{2na}\,{\bf u} \cdot {\bf e} + {\bf
  \Omega}_A \cdot {\bf k}=C_1,\ \ \ \mbox{and} \ \ \ 
-\frac{3A_0}{2na}\,{\bf u} \cdot {\bf k} + {\bf
  \Omega}_A \cdot {\bf e}=C_2. \label{constants}
\end{equation}
To find the excitation under the precessing acceleration, we introduce the two
vectors ${\bf k}_\pm={\bf k}\pm {\bf e}$ which decouple the equations of
motion and yield:
\begin{equation}
\frac{{\rm d}{\bf k}_\pm}{{\rm d}t}= \pm \frac{3A_0}{2na}\,  {\bf u}\times
{\bf k}_\pm,\ \ \mbox{and} \ \ 
{\bf k}_\pm \cdot \left[{\bf
  \Omega}_A \mp \frac{3A_0}{2na}
  \,{\bf u} \right]=C_\pm \label{kplusminusdot}
\end{equation}
where $C_\pm=C_1\pm C_2$. Note that $|{\bf k}_\pm|=1$ is conserved, a
consequence of ${\bf k}\cdot {\bf e}=0 $ and ${\bf e}^2+{\bf k}^2=1.$ We seek
a solution for the vectors ${\bf k}_\pm$ through their projection on the basis
made up of ${\bf \Omega}_A$, ${\bf u}$, and ${\bf \Omega}_A\times {\bf
u}$. Simple algebra shows that $ {\bf \Omega}_A \cdot {\bf k}_\pm$ satisfy the
equations:
\begin{eqnarray}
 \frac{{\rm d}^2 ({\bf \Omega}_A \cdot {\bf k}_\pm)}{{\rm d}t^2}
+\left[n^2_A/4+\Omega_A^2\mp n_A({\bf \Omega}_A\cdot {\bf u})\right]
({\bf \Omega}_A \cdot {\bf k}_\pm) =\nonumber &&\\ \left[\Omega_A^2\mp n_A({\bf \Omega}_A\cdot {\bf u})/2\right] C_\pm,&& \label{okdot}
\end{eqnarray}
which are those of  two harmonic oscillators of  frequencies $n_\pm$:
\begin{equation}
n_\pm^2=n^2_A/4+\Omega_A^2\mp n_A({\bf \Omega}_A\cdot {\bf u})
=({\bf \Omega}_A\mp{\bf \Omega}_k)^2.
\label{nomega}
\end{equation}
where ${\bf \Omega}_k=3A_0{\bf u}/2na$ is the instantaneous rotation vector of
${\bf k}_+$. The solutions of the previous equations are:
\begin{equation}
{\bf \Omega}_A \cdot {\bf k}_\pm =\frac{\left[2\Omega_A^2\mp n_A({\bf \Omega}_A\cdot {\bf u})\right]C_\pm}{2n_\pm^2} + K_\pm \cos \left(n_\pm t +\phi_\pm\right). \label{ok}
\end{equation}
where $K_\pm$ and $\phi_\pm$ are constants to be determined from the initial
conditions.  The projections on ${\bf u}$ are obtained from the constants of
motion as:
\begin{equation}
\mp{\bf u} \cdot {\bf k}_\pm =\frac{\left[n_A\mp 2({\bf \Omega}_A\cdot {\bf u})\right] C_\pm }
{2n^2_\pm}- \frac{2K_\pm}{n_A} \cos \left(n_\pm\, t +\phi_\pm\right). \label{uk}
\end{equation}

The projections along ${\bf \Omega}_A\times {\bf u}$ are found by noting that
$({\bf \Omega}_A\times {\bf u}) \cdot {\bf k}_\pm= \pm 2 {\rm d}({\bf
\Omega}_A \cdot {\bf k}_\pm)/ {\rm d}t/n_A$ which lead to:
\begin{equation}
 ({\bf \Omega}_A\times {\bf u})  \cdot {\bf k}_\pm =\mp\frac{2n_\pm K_\pm
    }{n_A} \sin \left(n_\pm\, t +\phi_\pm\right).\label{ouk}
\end{equation}
For initially circular orbits that interest us, the integration constants are
given by:
\begin{eqnarray}
&&K_\pm \cos\phi_\pm=\frac{\left[n_A^2\mp 2n_A({\bf
 \Omega}_A\cdot {\bf u})\right] C_1}{4 n_\pm^2
 }\nonumber \\ && \ \ \ \ \ \ \ \ 
\ \ \ \ \ \  \ \ \ \ \  \ \
+\frac{\left[2n_A({\bf
 \Omega}_A\cdot {\bf u})\mp 4\Omega_A^2\right]C_2}{4 n_\pm^2
 },  \\&& K_\pm \sin\phi_\pm=\mp\frac{n_AC_0}{2n_\pm}.
\end{eqnarray} 
where $C_0$ denotes the initial value of $({\bf \Omega}_A\times {\bf u}) \cdot
{\bf k}_\pm$. The eccentricity and inclination expressions are obtained from
$e=\sqrt{(1-{\bf k}_{+}\cdot {\bf k}_{-})/2}$ and $\cos I={\bf
\Omega}_A\cdot({\bf k}_{+}+{\bf k}_{-})/2\Omega_A\sqrt{1-e^2}$ (note that the
components of ${\bf k}_\pm$ are not given in an orthogonal basis; to recover
${\bf k}_{+}\cdot {\bf k}_{-}$ easily, an additional step is needed and
consists of writing ${\bf u}$ in terms of its invariant part along ${\bf
\Omega}_A$ and its precessing component along the unit vector ${\bf u}_A$ as
${\bf u}=\sqrt{1-({\bf \Omega}_A\cdot{\bf u})^2} {\bf u}_A+({\bf
\Omega}_A\cdot{\bf u} )\, \Omega_A^{-2}{\bf \Omega}_A$ and substituting it
into equations (\ref{uk}) and (\ref{ouk})). This completes the solution of the
excitation by a precessing acceleration.

The eccentricity excitation differs from that without precession in three
ways: (i) the excitation amplitude depends on the ratio $n_A/2\Omega_A$, (ii)
the motion involves two fundamental frequencies $n_\pm$ if ${\bf\Omega}_A$ and
${\bf u}$ are not orthogonal, and (iii) large eccentricity excitation becomes
accessible at all initial relative inclinations through the resonance
$n_A=2\Omega_A$ where the frequencies $n_\pm$ in the denominators of the
amplitudes of ${\bf k}_\pm$ become small.

In the following, we apply these findings to the case of a precessing jet
whose rotation vector ${\bf \Omega}_A$ is parallel to the initial angular
momentum vector or equivalently the initial vector ${\bf k}$. The constants of
integration are given as: $C_0=0$, $C_1=\Omega_A$ and $C_2=-n_A \cos (\alpha)
/2$ where $\alpha$ is the angle between ${\bf u}$ and ${\bf
\Omega}_A$. Denoting by $p$ the frequency ratio $n_A/2\Omega_A$ and using the
solution derived above, we find:
\begin{eqnarray}
e^2&=&\frac{p^2\sin^2\alpha}{4 \nu_{+}^2\nu_{-}^2}\left[2  (3 + p^2)
-4 (1+p \cos\alpha  )\ \cos \nu_{+}t 
-4 ( 1- p \cos\alpha)\  \cos \nu_{-}t\right.\nonumber\\
&& +( 1-p^2+\nu_{+}\nu_{-}) \cos (\nu_{+}-\nu_{-})t
\left.+(1-p^2-\nu_{+}\nu_{-})\cos (\nu_{+}+\nu_{-})t\right]\label{e-jet}\\
\cos I&=&\frac{1}{2
  \nu_{+}^2\nu_{-}^2\sqrt{1-e^2}}\left[(p^4-p^2+2+p^2[p^2-3]\cos 2\alpha)\nonumber\right.\\
&& \left.+p^2\sin\alpha^2(p^2+1+2 p\cos\alpha)\, \cos\nu_{+}t + p^2\sin\alpha^2(p^2+1-2 p\cos\alpha)\, \cos\nu_{-}t \right]
 \label{i-jet}
\end{eqnarray}
where $\nu_{\pm}^2=(n_\pm/2\Omega_A)^2=p^2+1\mp 2 p\cos\alpha$ and $t$ is
normalized by $\Omega_A$; for definiteness we take $A_0>0$ in what follows.

\begin{figure}\epsscale{.7}
\plotone{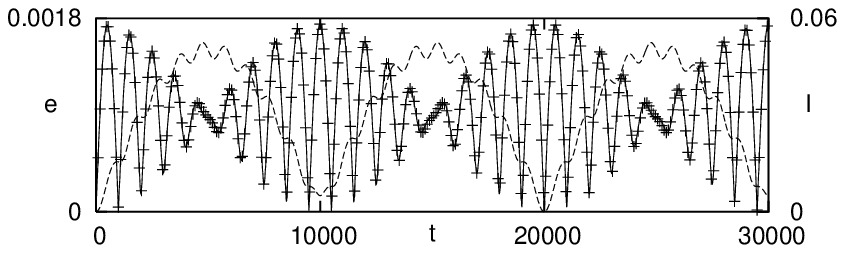}\\
\plotone{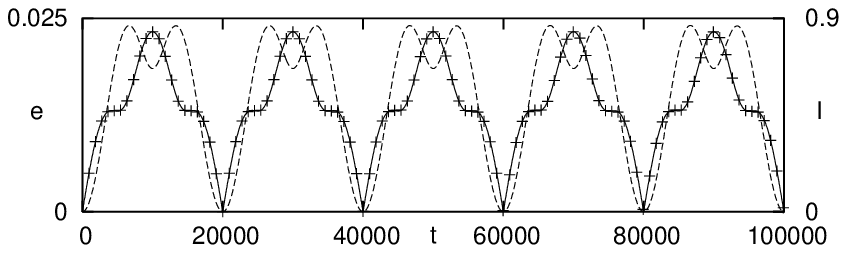}\\
\plotone{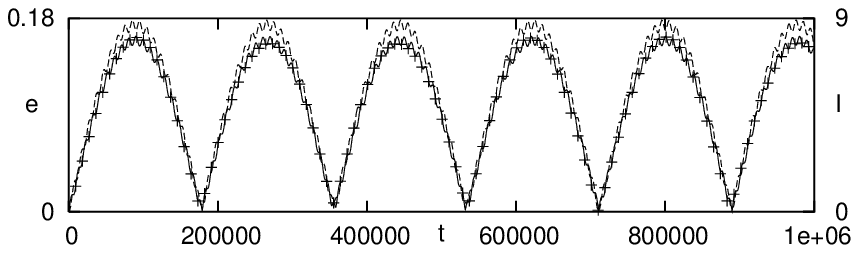}\\
\plotone{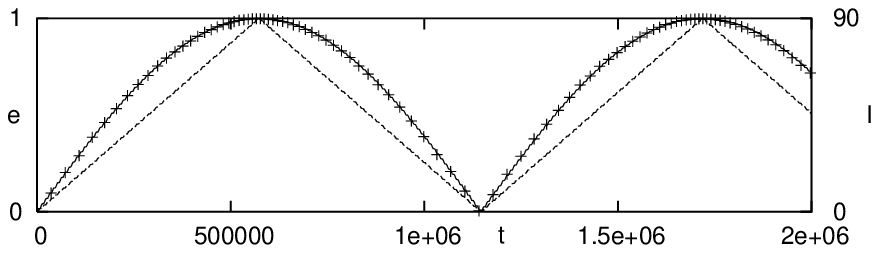}\\
\plotone{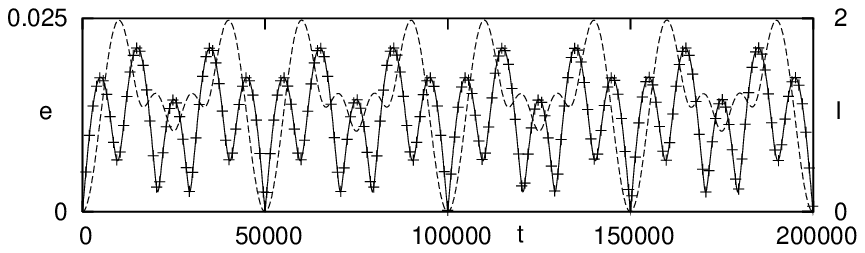}
\caption{Excitation of the eccentricity (solid) and inclination (dashed) by a
         nearly perpendicular precessing jet with an angle $\alpha=1^\circ$.
         The companion's orbit is located at $a=1$ AU and evolves from a
         circular orbit in a plane orthogonal to the jet's precession
         axis. The acceleration is $A_0=2\times 10^{-10}{\rm km\,s}^{-2}$
         yielding an excitation time of $2\pi/n_A=10^4$ years. From top to
         bottom, the panels were obtained from equations
         (\ref{e-jet}--\ref{i-jet}) with the frequency ratios, $p$: 0.05, 0.5,
         0.9, 1 and 5 -- the precession period is $2p\times 10^4$ years. The
         symbols correspond to the numerical integration of the full equations
         of motion (\ref{motion}) and show that the agreement with the
         averaged analytical solution is perfect.}
\label{fig-4}
\end{figure}

\begin{figure}\epsscale{.7}
\plotone{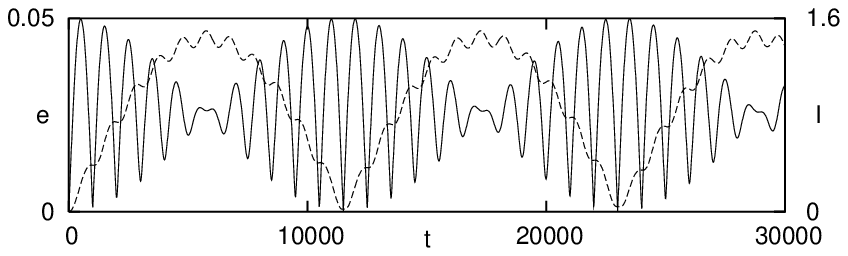}\\
\plotone{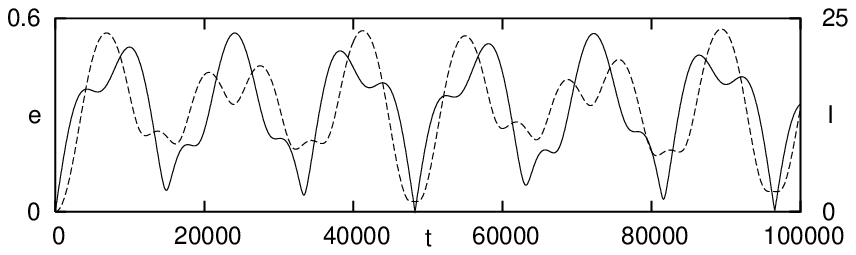}\\
\plotone{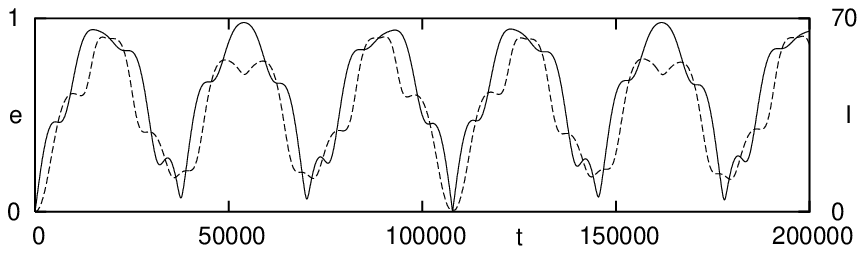}\\
\plotone{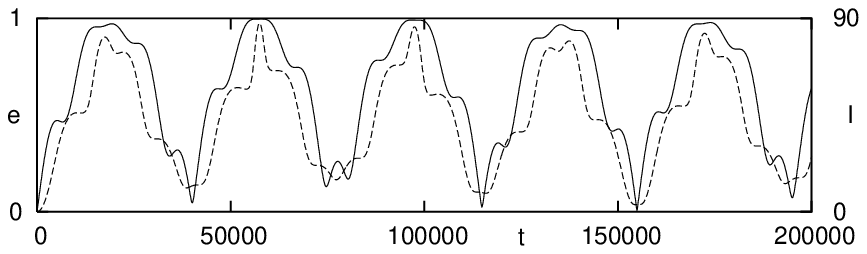}\\
\plotone{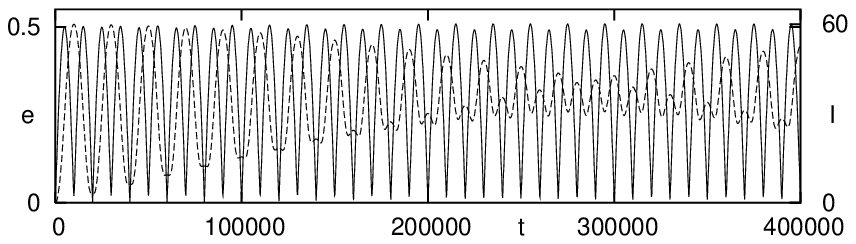}
\caption{Same as Figure (\ref{fig-4}) for a jet angle $\alpha=30^\circ$. From
         top to bottom, the panels correspond to the frequency ratios, $p$:
         0.05, 0.5, 0.9, 1 and 50.}
\label{fig-5}
\end{figure}

The location where the frequency match, $p=1$, occurs defines the nominal
resonant semi-major axis $a_{\rm res}=G(M+m)(2\Omega_A/3A_0)^2$. The frequency
ratio can be written as $p=\sqrt{a/a_{\rm res}}$.  We consider the differences
introduced by precession in the three regions: far inside resonance ($p\ll
1$), far outside resonance ($p\gg 1$) and the resonance region $p=1$. Far
inside resonance ($p\ll1$), the jet precesses faster than the eccentricity
excitation leading to a reduction of the eccentricity amplitude from
$\sin\alpha$ to $2p\sin\alpha$ as $e=p|\sin\alpha|\left[(3+\cos
[2p\cos(\alpha)t]-4 \cos[t]\cos[p\cos(\alpha) t])/2\right]^{1/2}$. Far outside
resonance ($p\gg 1$), the jet's precession is slow compared to the
eccentricity excitation so that the latter is described by the expressions
given in section 3 for a constant-direction acceleration; the inclination,
however, is not.  The slow precession causes a modulation of the inclination
oscillation between $0$ and $\sin 2\alpha$ as $\cos
I=[\sin^2\alpha\cos(pt)\cos(\cos[\alpha]t)+\cos^2\alpha]/[1-\sin^2\alpha
\sin^2(pt)]^{1/2}$.  In the resonance region, the proximity of $p$ to unity
increases the denominators of the eccentricity expression (\ref{e-jet}) which
leads to eccentricities close to unity. The width of the region around
resonance where large eccentricity values are reached increases with the jet
angle $\alpha$. These features are illustrated in Figures (\ref{fig-4}) and
(\ref{fig-5}) where we plot the expressions (\ref{e-jet}) and (\ref{i-jet})
for two jet angles $\alpha=1^\circ$ and $30^\circ$, an excitation period
$2\pi/n_A=10^4$ years, and the four values of $p$: 0.05, 0.5, 0.9, 1 and
50. Figure (\ref{fig-4}) also includes the result of the integration of the
full equation of motion (\ref{motion}) and shows that the solution equations
are indistinguishable from the unaveraged numerical solution. This agreement
results from the fact that the acceleration is independent of the relative
motion and the relative velocity. The figures show that the resonance region
does not extend far around $p=1$ for a jet angle $\alpha=1^\circ$, as the
maximal amplitude for $p=0.9$ is 0.18 while for $\alpha=30^\circ$ the
resonance region is much wider. We note that consistency in the use of the
secular potential requires that $p$ is not too small or equivalently that
$\Omega_A\ll n$. Numerical integrations of the full equations of motion show
that the eccentricity and inclination expressions can be used as long as
$\Omega_A<0.1 \,n$. Moreover, as the eccentricity excitation time is $n_A$, no
resonant forcing occurs when $\Omega_A=n$ in the secular region ($n_A\ll n$).

Precession-driven resonant excitation therefore provides a possible way to
raise the eccentricity even in low jet angle systems.  If a companion happens
to be at or cross the excitation region, because of disk-driven migration, not
only that the eccentricity will grow but the planet leaves the disk as its
inclination is excited in phase with the eccentricity. The tidal interaction
of the disk with the planet will not prevent the eccentricity excitation if
the precession period of the jet, $10^2$ to $10^4$ years \citep{b55,b54,b51},
is shorter than the viscous time of the disk, $10^5$ to $10^6$ years
\citep{b1}.  Such a situation offers a possible prospect for stopping
migration while exciting the eccentricity provided that (i) the disk is stable
to the perturbing acceleration and that (ii) the planet's migration time is
longer than the excitation time.  These results are discussed further after we
constrain the magnitude of the perturbing acceleration in section 8.

\section{Effect of radial migration on eccentricity excitation}
The changes in the orbital semi-major axis $a$ of the companion have so far
been neglected because in the secular excitation region where $n_A$ is smaller
than the mean motion $n$, a constant acceleration will produce small periodic
oscillations of the semi-major axis with frequency $n$. It is possible that
radial migration occurs indirectly even for an acceleration that is weakly
dependent on the local orbital dynamics. Consider for instance the case where
acceleration is caused by mass outflow from the star-disk system.  The
planet's orbital revolution is determined by the matter content inside its
orbit. If the total mass of that content varies over timescales larger than
the orbital period the conservation of angular momentum leads to the radial
migration of the companion but does not affect its eccentricity \citep{b10}
leaving initially circular orbits invariant.  This conservation leads to the
relation $Ma=M_0a_0$ which in the case of mass loss induces outward
migration. To estimate the related migration, we note that the disk's mass
inside the planetary companion's orbit is at most several percent of $M_0$. If
all of this mass is ejected from around the star, the companion's semi-major
axis expands by a corresponding several percent showing that in this case
migration is not significant.

Disk-companion interactions usually yields an inward radial migration.  To
find the effect of migration on the eccentricity excitation, we simulate the
disk-companion interaction by the addition of a Stokes-type drag $-k{\bf v}$
to the equations of motion (\ref{motion}) where $k$ may be a function of time
whose characteristic timescale is larger than the orbital period.  Simple
algebra shows that in the absence of external acceleration, the drag term
conserves the modified angular momentum $\eta \, {\bf h}$ where $\eta=\exp\int
k {\rm d}t$. In terms of osculating orbital elements, the previous relation
implies that $a \,(1-e^2) \, \eta^2$ is conserved showing that the drag term
leads to orbital migration of the companion with respect to the star. The time
dependence of $k$ can be interpreted as the mathematical representation of the
decay law of $a$; for instance, if $a=a_0(t/\tau_a+1)^{-\alpha}$, then
$k=-\alpha(t/\tau_a+1)^{-1}/2\tau_a$ with $n\tau_a\ll 1$ where $\tau_a$ is the
migration timescale that depends on the parameters that regulate angular
momentum transfer between the companion and the gas disk. A constant $k$
implies an exponential decay of $a$.  When the migration time is larger than
the companion's orbital period, which is usually the case, the average effect
of this drag term conserves the orbit's eccentricity and the planet stays in
its initial circular orbit while migrating with respect to the star.

\begin{figure}\epsscale{.7}
\plotone{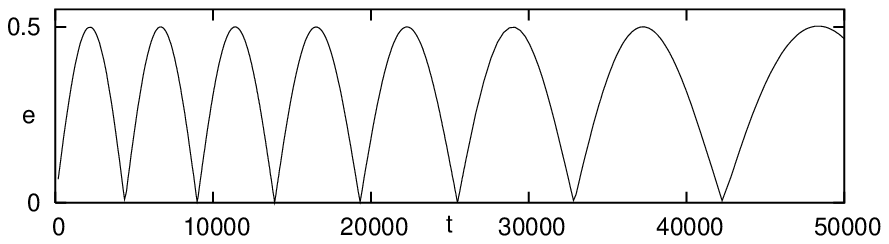}\\
\plotone{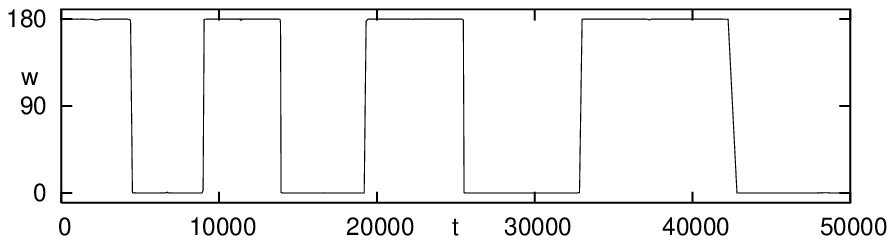}\\
\plotone{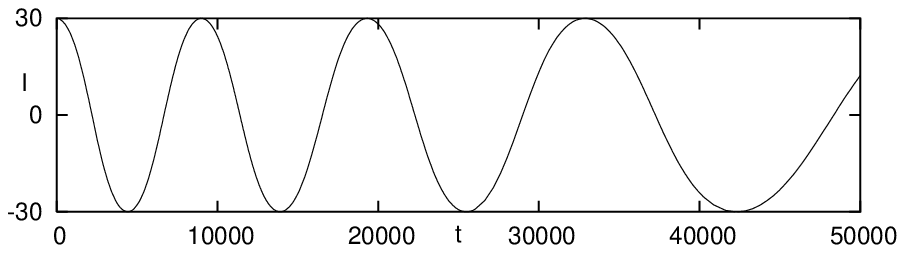}\\
\plotone{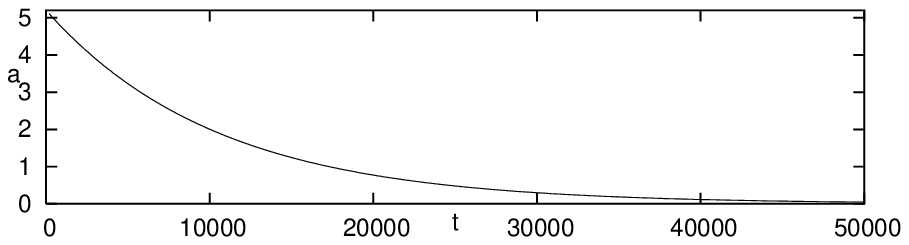}
\caption{Effect of radial migration on eccentricity excitation.  In this
         numerical integration of (\ref{motion}) with a constant direction
         acceleration of magnitude $A_0=2\times 10^{-10}{\rm km\,s}^{-2}$ and
         inclination $I_0=30^\circ$, a Stokes drag term, $-k{\bf v}$ with
         $k^{-1}=36\,500$ years, was added to the equations of motion. From
         top to bottom, the panels show the eccentricity, argument of
         pericenter, the inclination and the semi-major axis. The companion is
         located at 5.2 AU on an initially circular orbit.  The initial
         excitation period is 4400 years. The slow migration with respect to
         the orbital motion affects the eccentricity evolution only through
         the oscillation period.}
\label{fig-6}
\end{figure}

Radial
migration would therefore affect the eccentricity excitation only through the
variation of the excitation frequency $n_A$ which is where the semi-major axis
enters the excitation mechanism. This is illustrated in Figure (\ref{fig-6})
where we show an example of the excitation of eccentricity during migration.
For multiplanet systems, the eccentricity excitation can be affected by the
migration significantly because the acceleration has to compete with mutual
planetary perturbations that tend to precess the planetary orbits at rates
that may be faster than the acceleration's excitation frequency.

\section{Keplerian boundary and sudden excitation}
\begin{figure}
\epsscale{.36}
\plotone{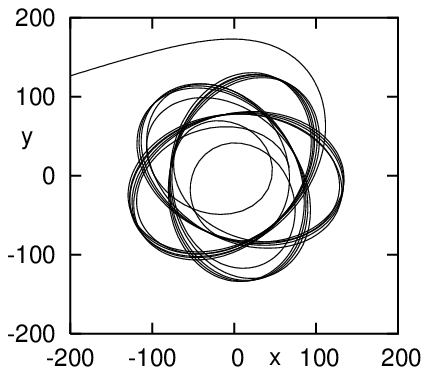}
\plotone{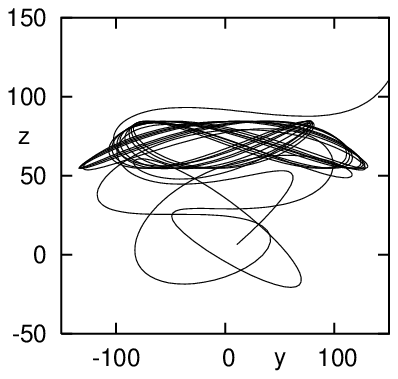}
\plotone{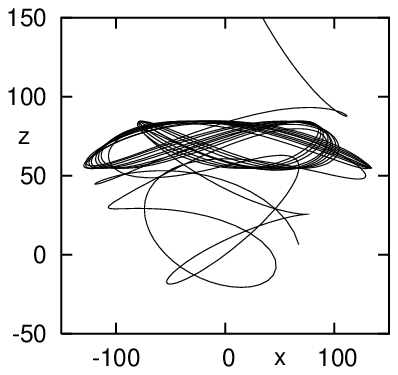}
\caption{Escape of a companion located near the keplerian boundary of a
         conservative constant-direction acceleration with $I_0=30^\circ$. The
         distances are given in AU. The boundary's semi-major axis is $a_{\rm
         kplr}=100$\,AU. Note how the companion hovers above the star before
         escaping. The corresponding orbital elements are shown in Figure
         (\ref{fig-8}).  }
\label{fig-7}
\end{figure}

\begin{figure}
\epsscale{.7}
\plotone{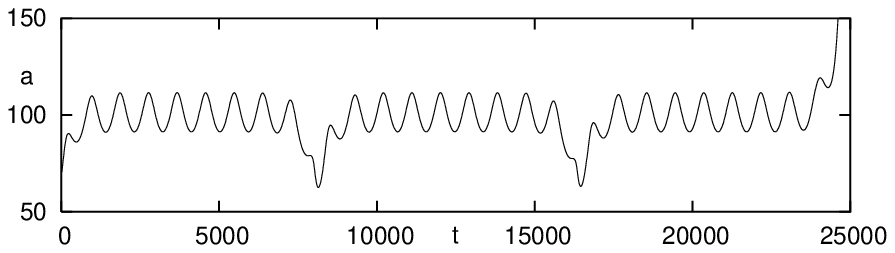}\\
\plotone{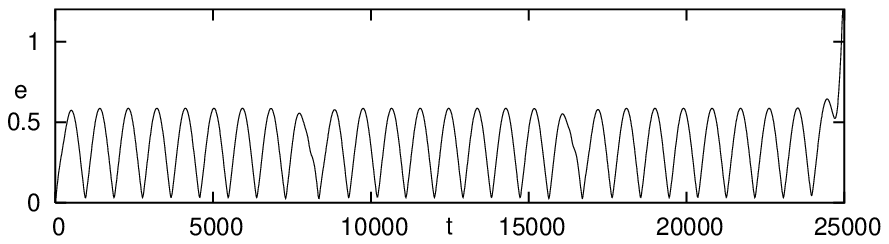}\\
\plotone{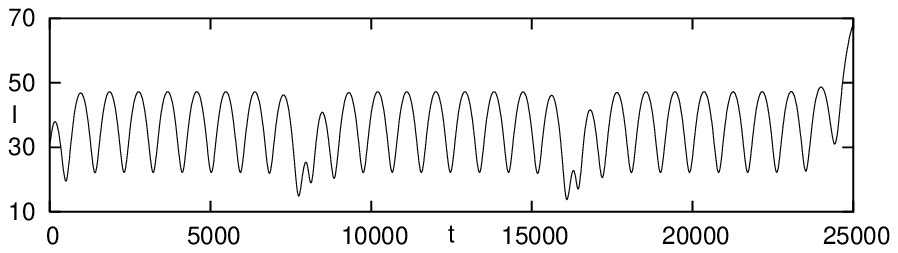}
\caption{Escape of a companion located near the keplerian boundary of a
         conservative constant-direction acceleration. The boundary's
         semi-major axis is $a_{\rm kplr}=100$\,AU. The companion is initially
         on a circular orbit of semi-major axis 68.5\,AU with $I_0=30^\circ$.
         The semi-major axis experiences large amplitude librations about
         $a_{\rm kplr}$. The eccentricity amplitude is larger than $\sin I_0$
         and the inclination evolution differs from that inside the keplerian
         region. }
\label{fig-8}
\end{figure}

In order to be stable, a companion's orbit needs to receive greater
acceleration from the star than from the perturbation. This condition
delineates the keplerian region around a star as that for which $G
(M+m)/r^2>|A|$. The natural limit for small orbital perturbations is located
closer to the star where the frequency $n_A$ becomes comparable to the local
mean motion $n$ of the companion.  Near this limit, the forced periodic
oscillations of the semi-major axis $a$ are reinforced by the eccentricity and
acquire large amplitudes. Denoting by $a_{\rm kplr}$ the semi-major axis of
the keplerian boundary where $n_A=n$, we obtain an expression for the
magnitude of the perturbing acceleration as:
\begin{equation}
|A_0|\simeq 
2\times 10^{-12}\left(\frac{M+m}{M_\odot}\right)\,\left(\frac{10^3\, {\rm AU}}{a_{\rm
 kplr}}\right)^2\ {\rm km}\, {\rm s}^{-2} \label{stability1}
\end{equation}
which for $a_{\rm kplr}=10^3$\,AU is an order of magnitude larger than the
acceleration of the solar system in the Galaxy, the match occurs near $a_{\rm
kplr}= 3000$ AU. The excitation period of this acceleration, $T_A=2\pi/n_A$,
is given as:
\begin{equation}
T_A\simeq
10^{6}\left(\frac{M+m}{M_\odot}\right)^\frac{1}{2}\,
      \left(\frac{a_{\rm kplr}}{10^3\, {\rm AU}}\right)^2\,
      \left(\frac{1\, {\rm AU}}{a}\right)^\frac{1}{2}\
   {\rm years} \label{TA}.
\end{equation}
When the boundary $a_{\rm kplr}$ of a given acceleration is located beyond
$10^4$ AU, the Galactic potential's tide becomes important (Heisler and
Tremaine 1986) and must be included to determine the limits of the
boundary. In this work, we will be concerned with smaller values of $a_{\rm
kplr}$. To illustrate the motion near the keplerian boundary, we show an
example of an escape orbit of a conservative constant-direction acceleration
with $a_{\rm kplr}=10^2$\,AU and an inclination $I_0=30^\circ$ (Figures
\ref{fig-7} and \ref{fig-8}). The orbit's initial semi-major axis is 68.5
AU. The characteristics of motion are not strictly keplerian as the companion
hovers above the star. The semi-major axis shows periodic oscillations around
100 AU with a significant amplitude. These characteristics depend to a certain
extent on the direction of acceleration. A study of the types of motion near
the boundary is interesting but not central to the problem of the eccentricity
excitation.  We remark that the escape orbits offer an interesting way to
expel planets from around their parent stars or equivalently to disrupt a
binary stellar system. If a companion is formed near the keplerian boundary or
is pushed out to it by a possibly remaining inner disk that followed
photo-evaporation \citep{b6,b31}, it could become unbound. Systems where the
acceleration keeps an approximately constant direction during eccentricity
excitation loose companions without a risk of catastrophic encounters with the
star if the initial direction of acceleration does not lie in the companion's
orbital plane such as that of Figure (\ref{fig-8}).

We remark that the value of $a_{\rm kplr}$ corresponds to the innermost
location of the stability boundary because the acceleration's decay decreases
$n_A$ and pushes the stability boundary outward. The acceleration's finite
duration may extend the keplerian boundary depending on the ratio of $\tau$ to
the excitation time at the keplerian boundary of the conservative problem
$T_A(a_{\rm kplr})/2$. When $\tau\geq T_A(a_{\rm kplr})/2$, orbits beyond
$a_{\rm kplr}$ have enough time to acquire sufficient momentum to escape the
gravitational pull of the star.  When $\tau\leq T_A(a_{\rm kplr})/2$, the
stability region extends beyond  $a_{\rm kplr}$ and its 
boundary is given by the semi-major axis where $\tau\simeq
T_A(a_\infty)/2$ which is larger than $a_{\rm kplr}$ since $T_A$ is a
decreasing function of the semi-major axis $a$. Using the expression of $T_A$
and the residual velocity $V$, $a_\infty\simeq G(M+m)V^{-2}$, the location
where the keplerian velocity matches $V$.

We now determine the features of orbital excitation in the keplerian boundary
for an acceleration whose duration, $\tau$, is smaller than the excitation
time at the keplerian boundary.  We consider the outermost orbits in the
sudden excitation region for which the excitation time is very small compared
to the orbital time. To these orbits, the sudden excitation imparts a near
instantaneous velocity ${\bf V}$.  The changes in the orbital elements are
found by expressing three conservation relations. First is the conservation of
the potential energy as the position ${\bf x}$ is left invariant during the
excitation. This is written as:
\begin{equation}
\frac{G(M+m)}{|{\bf x}|}=\frac{G(M+m)}{2a_{\rm i}}+\frac{1}{2} \,
 v^2_{\rm i}
=\frac{G(M+m)}{2a_{\rm f}}+\frac{1}{2} \,
 v^2_{\rm f},
\label{sudden1}
\end{equation}
where $a_{\rm i}$, $a_{\rm f}$, $v_{\rm i}$ and $v_{\rm f}$ are the initial
and final semi-major axes and velocities of the companion.  Eliminating the
potential energy term leads to:
\begin{equation}
\frac{G(M+m)}{a_{\rm f}}=
\frac{G(M+m)}{a_{\rm i}}-2{\bf  v}_{\rm i}\cdot {\bf V}
-V^2.
\label{sudden2}
\end{equation}
For an initially circular companion orbit in a coordinate system where the
$z$-axis is chosen along ${\bf V}$ (section 3), the previous equation becomes:
\begin{equation}
\frac{1}{a_{\rm f}}=\frac{1}{a_{\rm i}}-\frac{2 V\sin
  I_0\cos\theta}{\sqrt{G(M+m) a_{\rm i}}}  -\frac{V^2}{G(M+m)},
\label{sudden3}
\end{equation}
where $\theta$ is the longitude of the companion along its orbit, and $I_0$ is
the inclination of the orbital plane with respect to the direction of the
residual velocity ${\bf V}$. This equation determines, for a given ${\bf V}$,
the final semi-major axis $a_{\rm f}$ as a function of $a_{\rm i}$ and
$\theta$.  We note that $a_{\rm f}$ can be larger or smaller than $a_{\rm i}$
because of the inclination term in the energy equation (\ref{sudden3}).  This
leads to an inward or outward migration of the companion. Note that when ${\bf
V}$ is perpendicular to the initial orbital plane ($I_0=0$), migration is
always outward.

\begin{figure}
\epsscale{.5}
\plotone{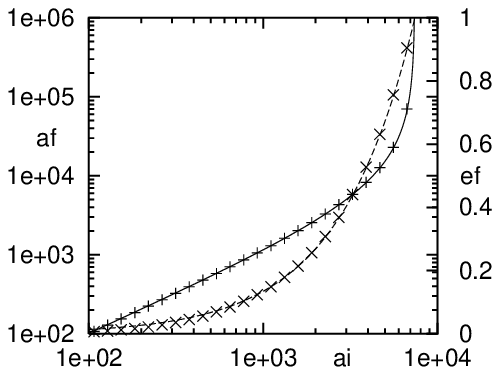}\plotone{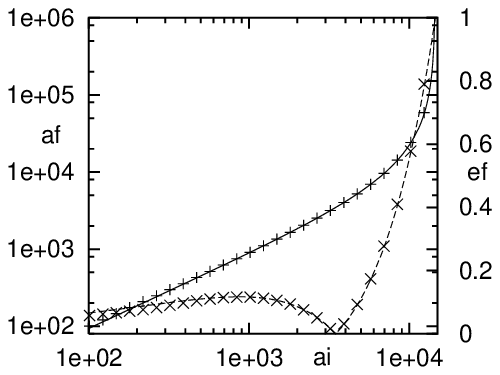}
\caption{Migration and eccentricity excitation near the keplerian boundary of
         a finite duration acceleration for two inclination values $I_0=0$
         (upper panel) and $I_0=20^\circ$ (lower panel).  In each panel, the
         final semi-major axis $a_{\rm f}$\,(AU) (solid) and the final
         eccentricity $e_{\rm f}$ (dashed) are shown as a function of the
         initial semi-major axis $a_{\rm i}$\,(AU) as given by equations
         (\ref{sudden3}) and the numerical integration of the full equations
         of motion (\ref{motion}) (symbols).  The parameters are:
         $V=0.35$\,km\,s$^{-1}$, $a_{\rm kplr}=300$\,AU, and $\tau=500$
         years. The inclined circular orbits were started at the descending
         node ($\theta=180^\circ$).  For $I_0=0^\circ$,
         $a_{\infty}=7\,341$\,AU and for $I_0=20^\circ$, $a_{\rm
         out}=3\,477$\,AU, $a_{\rm esc}=3\,797$\,AU,
         $a_{\infty}=14\,542$\,AU.}
\label{fig-9}
\end{figure}

\begin{figure}
\epsscale{.5}
\plotone{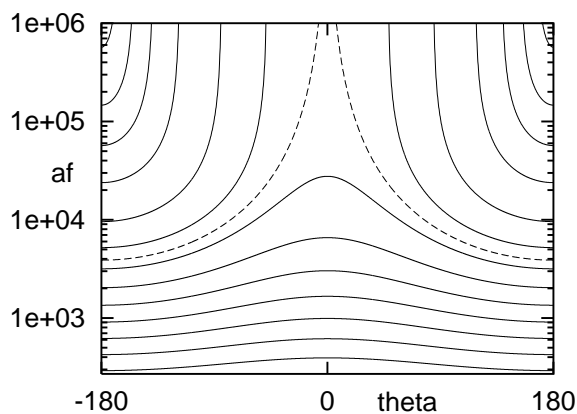}
\caption{Final semi-major axes some of the initially inclined orbits
        ($I_0=20^\circ$) of Figure (\ref{fig-9}) as a function of orbital
        longitude $\theta$.  The dashed curve corresponds an initial
        semi-major axis at $a_{\rm esc}$.}
\label{fig-10}
\end{figure}

The second relation is obtained from the conservation of the linear momentum,
${\bf v}_{\rm f}={\bf v}_{\rm i}+{\bf V}$. For initially circular orbits, the
projection of the velocity on the position vector yields: ${\bf x}\cdot {\bf
v}_{\rm f}={\bf x}\cdot {\bf V}$. In the cases that we consider below where
the companion is initially at the orbit's nodes, ${\bf x}\cdot {\bf v}_{\rm
f}=0$ implying the conservation of the pericenter distance $ a_{\rm
f}(1-e_{\rm f})=a_{\rm i}$ if migration is outward and the apocenter distance
$a_{\rm f}(1+e_{\rm f})=a_{\rm i}$ if migration is inward. This leads to the
following relation that is valid at the nodes:
\begin{equation}
e_{\rm f}=|1-a_{\rm i}/a_{\rm f}|.
\end{equation}
Elsewhere on the orbit, one needs to express the projections of the linear
momentum conservation equation in order to derive the final eccentricity,
longitude of pericenter and longitude of ascending node as functions of the
longitude $\theta$.  The third relation comes from the conservation of the
projection of the angular momentum on ${\bf V}$:
\begin{equation}
\sqrt{a_{\rm f}(1-e^2_{\rm f})}\, \cos I_{\rm f}=\sqrt{a_{\rm i}}\, \cos I_0,
\label{sudden4}
\end{equation}
where $I_{\rm f}$ is the final inclination of the companion. There are three
particular initial semi-major axes that characterize migration: first is
$a_{\rm esc}$, the semi-major axis beyond which the companion can escape at
certain longitudes. Second is $a_\infty$, the semi-major axis beyond which all
companions are lost.  Third is $a_{\rm out}$, the semi-major axis beyond which
only outward migration occurs.  Using equation (\ref{sudden3}), we find:
\begin{eqnarray}
a_{\rm out}&=&\frac{G(M+m)}{V^2}\left(2\sin I_0\right)^2,\\
a_{\rm esc} &=&\frac{G(M+m)}{V^2}\left(-\sin I_0 + \sqrt{1+\sin^2I_0}\right)^2,\\
a_\infty&=&\frac{G(M+m)}{V^2}\left( \sin I_0 + \sqrt{1+\sin^2I_0}\right)^2.
\label{sudden5}
\end{eqnarray}
Migration is inward in the longitude range around $\theta=180^\circ$ defined
by:
\begin{equation}
\theta_{\rm rev}^\pm=180^\circ\pm\cos^{-1}\left|\frac{\sqrt{a_{\rm i}}}{2V \sin I_0}\right|.
\label{sudden6}
\end{equation}
Between $a_{\rm esc}$ and $a_\infty$, escape occurs in the longitude range
around $\theta=0$ defined by:
\begin{equation}
\theta_{\rm esc}^\pm=\pm\cos^{-1}\left|\frac{a_{\rm i}^{-1}- V^2/G(M+m)}{2 V \sin I_0/\sqrt{G(M+m) a_{\rm i}}}\right|.
\end{equation}
Figure (\ref{fig-9}) shows the final semi-major axes and eccentricities at two
different inclinations $I_0=0$ and 20$^\circ$ for an acceleration
corresponding to $a_{\rm kplr}=300$\,AU and a duration $\tau=500$ years
resulting in a residual velocity $V=0.35$\,km\,s$^{-1}$. The companions were
started at the descending node for $I_0=20^\circ$ to illustrate inward and
outward migration. Also shown in the same figure are the results of the
integration of the full equations of motion (\ref{motion}) with the form
$A_1(t)$ confirming the validity of the analytic expressions in the outer
keplerian boundary. An example of migration with the form $A_2$ is used for
the eccentricity excitation of the $\upsilon$~Andromedae system in section
8.1. The dependence of migration on the orbital longitude is illustrated in
Figure (\ref{fig-10}) for the inclined orbits with $I_0=20^\circ$. The
migration and excitation near the keplerian boundary offers a possible way to
transport minor bodies in the solar system (section 8.2).  For planetary
companions, the radial migration that results from the interaction with the
disk (section 5) has to be added to the migration in the sudden excitation
region in order to ascertain the dynamics in the outer keplerian boundary.

The previous study of orbital stability points out similar constraints for the
host star that is accelerated in its motion within the Galaxy. Stellar jets
for instance modify the star's velocity but its galactic motion precludes that
$|A_0|$ be larger than that the galactic acceleration for an arbitrarily
extended time else the star is ejected. The acceleration's duration, $\tau$,
is therefore an important parameter that determines the star's orbital
stability.  A limit on $\tau$ can be set by requiring that the residual
velocity, $V$, that contributes to the stellar random motion be smaller than
the known velocity dispersion, $\left<v_g\right>$. The condition
$V\leq\left<v_g\right>$ yields:
\begin{equation}
\tau \leq 
10^5 
 \ \frac{3\times
10^{-12}\,{\rm km\,s}^{-2}}{A_0}\ \frac{\left<v_g\right>}{10\, {\rm km\, s}^{-1}} \, {\rm years},
\end{equation}
which in terms of the size of the keplerian boundary is:
\begin{equation}
\tau \lesssim 
10^5 
  \left(\frac{M_\odot}{M+m}\right)\,\left(\frac{a_{\rm
 kplr}}{10^3\, {\rm AU}}\right)^2\frac{\left<v_g\right>}{10\, {\rm km\, s}^{-1}} \, {\rm years},
\end{equation}
where we used $V\sim A_0\tau$. For a precessing acceleration, the duration is
extended by a factor $1/\cos\alpha$ where $\alpha$ is the jet angle. In
addition, the previous estimate is meaningful for a single acceleration event
and does not account for an arbitrary time evolution of the angle $\alpha$.

\section{Mutual planetary perturbations}
\begin{figure}\epsscale{.7}
\plotone{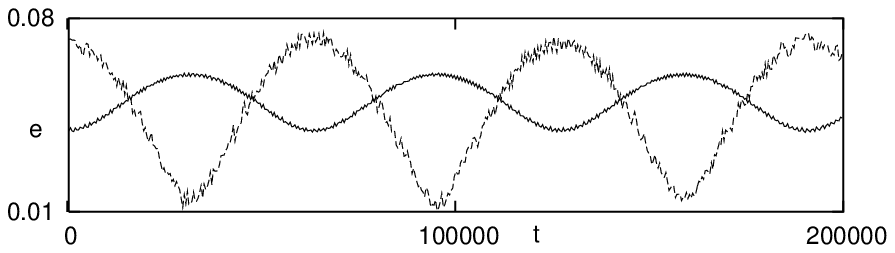}\\
\plotone{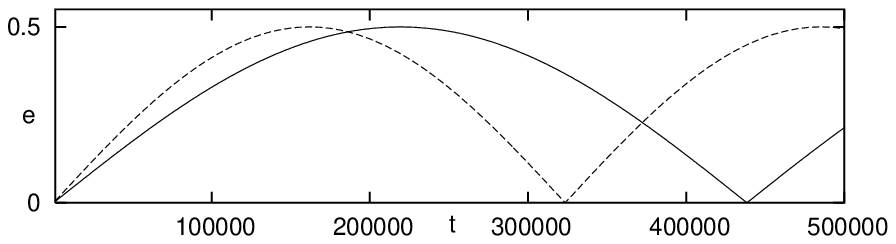}
\caption{The evolution of the eccentricity of Jupiter (solid) and Saturn
         (dashed). In the top panel mutual interactions are included but no
         external acceleration is present. The planets' initial eccentricities
         are $e_{\rm J}=0.04$ and $e_{\rm S}=0.07$. In the bottom panel, an
         external acceleration with $a_{\rm kplr}=10^3$\,AU acts on the two
         planets initially on circular coplanar orbits while mutual
         interactions are turned off. }
\label{fig-11}
\end{figure}

\begin{figure}\epsscale{.7}
\plotone{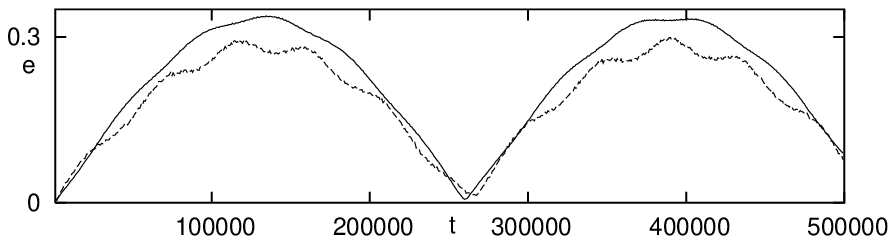}\\
\plotone{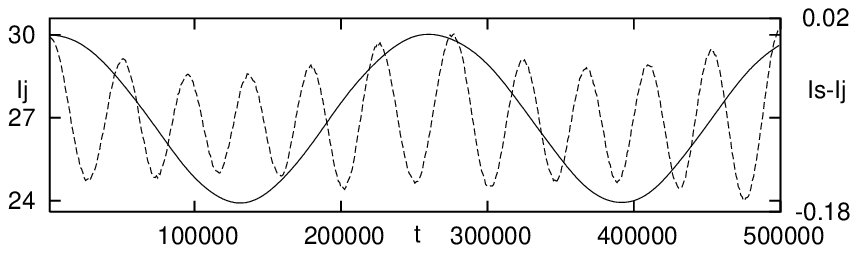}\\
\plotone{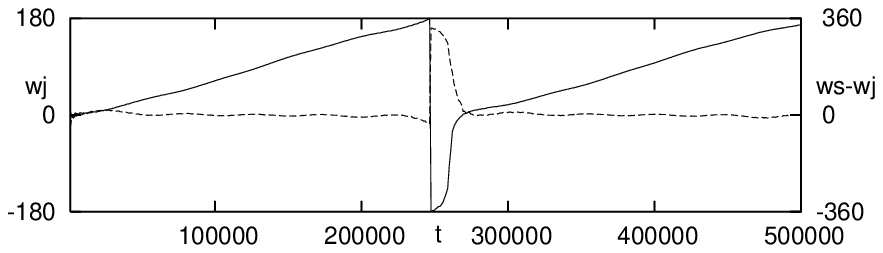}\\
\plotone{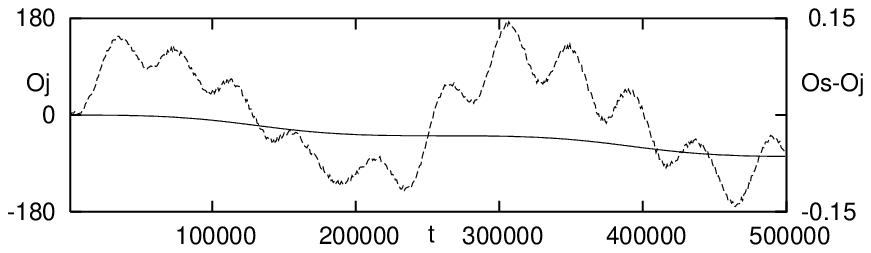}
\caption{The excitation of Jupiter's and Saturn's eccentricities and
  inclinations by a constant-direction acceleration with $a_{\rm kplr}=10^3$\,
 AU including the planets' mutual gravitational interactions.The planets were
  started with circular co-planar orbits.  From top to bottom, the panels show the evolution of the eccentricities,
Jupiter's inclination (solid) and the mutual inclination (dashed), 
Jupiter's argument of
  perihelion  (solid) and the mutual argument of perihelion (dashed), Jupiter's longitude of
  ascending node  (solid) and the mutual longitude of ascending node (dashed). 
 }
\label{fig-12}
\end{figure}

\begin{figure}\epsscale{.7}
\plotone{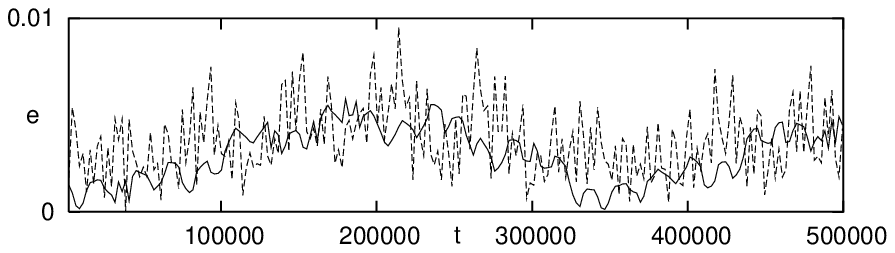}\\
\plotone{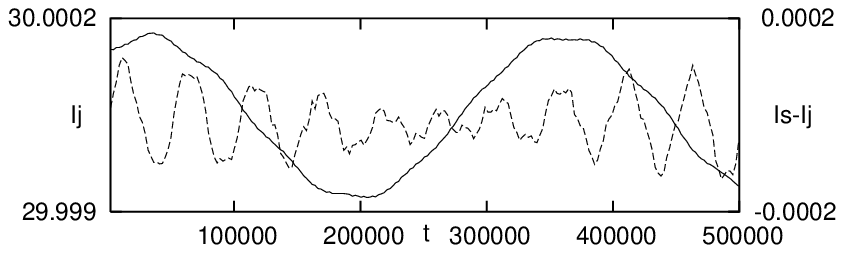}\\
\plotone{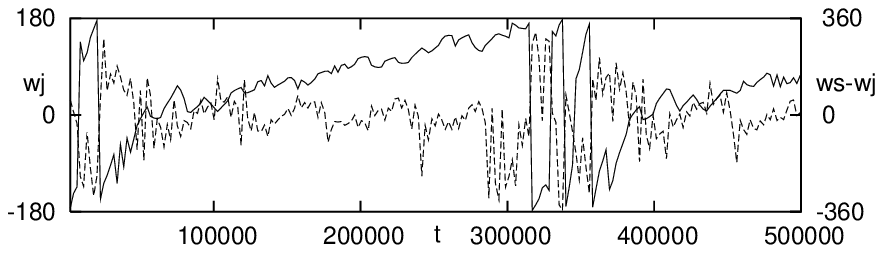}\\
\plotone{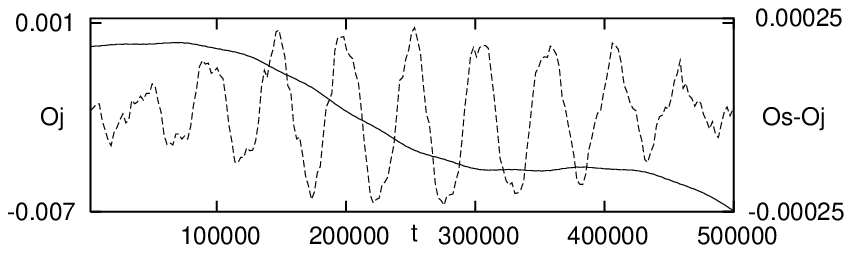}
\caption{The excitation of Jupiter's and Saturn's eccentricities and
  inclinations by a constant direction acceleration with $a_{\rm kplr}=10^4$\,
 AU including the planets' mutual gravitational interactions. The planets were
  started with circular co-planar orbits. From top to bottom, the panels show the evolution of the eccentricities,
Jupiter's inclination (solid) and the mutual inclination (dashed), 
Jupiter's argument of
  perihelion (solid) and the mutual argument of perihelion (dashed), Jupiter's
  longitude of
  ascending node (solid) and the mutual longitude of ascending node 
(dashed). 
 }
\label{fig-13}
\end{figure}

In section 4, we showed that the precession of the perturbing acceleration
affects significantly its ability to excite the companion's eccentricity. This
is also true if the acceleration does not precess but the system contains
multiple companions. In this case, mutual gravitational interactions cause the
eccentricity vectors to precess and possibly to be locked into resonance. We
expect for instance that if the precession rates are much faster than the
excitation frequencies, the maximum eccentricities will be reduced.  To
illustrate the effects of mutual interactions, we model the Jupiter-Saturn
system under the influence of a constant direction acceleration using a direct
integration of the equations of motion.  We assume that the planets formed in
initially coplanar circular orbits and that the planetary orbital plane is
inclined by 30$^\circ$ with respect to the direction of acceleration. The
semi-major axes and masses are taken as $a_{\rm J}=5.20$\,AU, $a_{\rm
S}=9.55$\,AU, $m_{\rm J}=9.55\times 10^{-4}M_\odot$ and $m_{\rm S}=2.86\times
10^{-4}M_\odot$. The unperturbed eigenfrequencies of the isolated
Jupiter-Saturn system are \citep{b17}: $9.6\times 10^{-4}\, ^\circ\, {\rm
year}^{-1}$ and $6\times 10^{-3}\,^\circ\, {\rm year}^{-1}$ for $e$ and
$-7\times 10^{-3}\,^\circ\, {\rm year}^{-1}$ for $I$ corresponding to the
eigenperiods $T_1=375\,000$ years, $T_2=60\,000$ years and $T_3=51\,000$ years
respectively. Note that the eccentricity amplitudes have a period of
$1/(1/60\,000-1/375\,000)\sim 71\,000$ years and that there is only one
inclination eigenmode due to the freedom of choosing a reference plane for
Jupiter's orbit. The evolution associated with mutual planetary perturbations
with no external acceleration is depicted in the top panel of Figure
(\ref{fig-11}) where we have ascribed to the two planets their current
eccentricities; the eigenfrequencies of the system are independent of $e$ and
$I$ if these are small. We assess the effect of mutual perturbation by
applying two different acceleration strengths, $A_0=2\times 10^{-12}\,{\rm
km\,s}^{-2}$ and $A_0=2\times 10^{-14}\,{\rm km\,s}^{-2}$, corresponding to
the two values $a_{\rm kplr}=10^3$\,AU and $a_{\rm kplr}=10^4$\,AU for the
location of the keplerian boundary. Taking $a_{\rm kplr}=10^3$\,AU yields
$T_{A\rm J}=4.4\times 10^5$ years, $T_{A\rm S}=3.2\times 10^5$ years; a
simulation of this configuration is shown in the bottom panel of Figure
(\ref{fig-11}) for which we removed mutual interactions. The case where both
mutual interactions and the perturbing acceleration are turned on is shown in
Figure (\ref{fig-12}).  The eccentricities of both planets are excited to 0.3
except that Saturn's is slightly smaller than Jupiter's.  This can be
explained by the fact that $T_{A\rm J}\sim T_1$ which implies a smaller
eccentricity reduction for Jupiter. In the case of Saturn, we have $T_{A\rm
S}\gg T_2$ which implies a stronger reduction than observed but the
gravitational interaction with a more massive Jupiter compensates that effect
and forces a larger eccentricity. The dominant eccentricity oscillation period
is modified to $2.6\times 10^5$ years.  For $a_{\rm kplr}=10^4$\,AU, $T_{A\rm
J}=4.4\times 10^6$ years, $T_{A\rm S}=3.2\times 10^6$ years. These timescales
are much larger that the unperturbed eigenperiods implying a strong reduction
of the excited eccentricities. Figure (\ref{fig-13}) shows that both
eccentricities are smaller than 0.01. The dominant eccentricity oscillation
period in this case is modified to $3.6\times 10^5$ years. We can find the
minimum strength of the constant-direction acceleration that can excite
Jupiter's eccentricity to its current value 0.05.  Numerical integrations
yield $a_{\rm kplr}=2600$\,AU corresponding to $A_0= 3\times
10^{-13}\,{\rm km\,s}^{-2}$.

We remark that during the eccentricity and inclination excitation, apsidal
alignment is maintained with the libration of $\Omega_{\rm S}-\Omega_{\rm J}$
and $\omega_{\rm S}-\omega_{\rm J}$ around zero. This is the result of the
forcing of the pericenters to be aligned with the direction of acceleration
that we encountered in the basic two-body problem of section 3. Note however
that the inclination oscillation about the direction of acceleration has been
largely suppressed. For $a_{\rm kplr}=10^3$\,AU (Figure \ref{fig-12}), the two
planets' inclinations oscillate around $26^\circ$ with an amplitude of
$3^\circ$ and a mutual inclination of $0.2^\circ$. We can use this feature as
another constraint on the initial strength of the constant direction force by
imposing the current mutual inclination of Jupiter and Saturn of $\sim
1^\circ$. Numerical integrations yield $a_{\rm kplr}=590$\,AU and
$A_0=5.7\times 10^{-12}\,{\rm km\,s}^{-2}$ with maximal eccentricities for
Jupiter and Saturn of $0.45$.

We can obtain an additional constraint on the perturbing constant-direction
acceleration by deriving its duration. We do this for the form $A_1$ with the
following simplified configuration: $a_{\rm kplr}=590$\,AU; the eccentricity
excitation occurs during the first oscillation cycle; disk driven radial
migration is neglected; and Jupiter and Saturn are assumed to have formed at
their current locations. In this case, equation (\ref{eoft}) leads to:
\begin{equation}
\tau \simeq \sin^{-1} \left( \frac{e_p}{\sin I_0}\right)\,  
\frac{T_e}{\pi},\label{decaytime}
\end{equation} 
where $e_p$ is the current eccentricity for Jupiter and $T_e\simeq 1.4\times
10^5$ years is the excitation period obtained with mutual perturbations. For
$e_p=0.05$, $\tau\sim 5000$\,years and the residual velocity $V\sim
1$\,km\,s$^{-1}$. The application of the model to the solar system is further
discussed in section 8.2. We finally comment that some of the dynamical
features that we have shown in this section can be reproduced analytically by
applying the Laplace-Lagrange secular perturbation theory at small $e$ and
$I$, and combining its second order secular gravitational potential with the
linearized acceleration potential (\ref{secpot}) (i.e. substituting $I$ for
sin $I$ in that expression). This would yield the linearized eigenfrequencies
and eigenvectors of the system. We do not follow this approach because we are
interested in large eccentricities and the possible effects of mean motion
resonances and close encounters.

\section{Discussion}
The eccentricity excitation theory that we have presented is based on the fact
that if a star-companion system is subjected to a relative acceleration that
is weakly dependent on the local dynamics, large eccentricities can be
achieved. The theory has the following parameters:
\begin{enumerate}
\item The magnitude of acceleration, $A$, which determines the excitation
      frequency of a star-companion system, and the extent of the keplerian
      region around the main star.
\item The direction of acceleration, ${\bf u}$, (unit vector) with respect to
      the initial orbital plane which determines the maximum eccentricity of a
      star-companion system as the projection of ${\bf u}$ on the orbital
      plane.
\item The rotation vector, ${\bf\Omega}_A$, which describes the acceleration's
      precession whose effect on the eccentricity amplitude and resonant
      excitation is described in section 4.
\item The duration of acceleration, $\tau$, which for a given eccentricity
      amplitude (larger than required) permits the selection of the final
      eccentricity value.
\end{enumerate}
When these parameters are specified, the final orbital state of a planetary or
stellar companion can be selected. The orbital configurations of star systems
with a single companion are easier to explain than those with multiple
companions because the absence of additional perturbers relaxes the
constraints on the acceleration's magnitude or equivalently the size of its
keplerian boundary $a_{\rm kplr}$. Since stellar jets and star-disk winds are
ubiquitous features of star formation and since we have established that they
can be dynamically responsible for the eccentricity excitation, the known
similarities and differences of the eccentricity distributions of planetary
companions and spectroscopic binaries can be attributed to the similarities
and differences of the physical environments that give rise to the
accelerating mass loss processes and not directly to the companions' formation
or their dynamical interaction with the gas disk. In this regard, it would be
interesting to be able to associate, for stars of a similar spectral type,
similar residual velocities that reproduce the observed companion
eccentricities.

The excitation theory is built on a minimal assumption that the perturbing
acceleration is explicitly independent of the local dynamics --the basic
excitation mechanism (section 3) relies on a constant acceleration that is
applied for a finite duration. As a result, multiplanet systems as well as
planetary systems of binary stars put strong constraints on the acceleration's
parameters. Once these parameters are specified, a single relative
acceleration is applied to all companions, planetary and stellar alike, and
sets the fundamental excitation frequencies and amplitudes of the
system. These frequencies are influenced by the mutual perturbations of the
companions as shown in section 7. In particular, the larger the companions'
number, the more constrained the acceleration's parameters.  In section 8.1,
we illustrate this situation with the $\upsilon$ Andromedae binary system
which contains multiple planets and a distant stellar companion whose observed
location strongly constrains the acceleration's strength. The application of
this theory to the known sample of multiplanet systems is beyond the scope of
this paper but it is ultimately the best way to ascertain whether the
acceleration mechanism is responsible for the eccentricities of extrasolar
planets. Two effects need to receive particular attention: the companions'
possible radial migration which changes the fundamental excitation frequencies
as seen in section 5, and the companion's secular perturbations which
circulate the eccentricities. Radial migration can be prescribed by the
standard disk-planet interaction theories \citep{b25}. The effect of secular
perturbations may be disentangled by determining for a given multiplanet
system the amplitudes of the secular modes from observed eccentricities and
seeking the acceleration parameters that reproduce the amplitudes of the
secular modes and not the observed eccentricities. This procedure would be
useful for similar mass planets because the instantaneous eccentricities can
be different from the individual secular mode amplitudes. The combination of
both migration and secular perturbations relaxes some of the constraints on
the model because of the additional freedom in the choice of the frequencies
of the secular modes. Such an analysis may help elucidate the
eccentricity-mass correlations of extra-solar planets. In the following, we
further illustrate how to constrain the acceleration mechanism in the case of
the $\upsilon$~Andromedae system (section 8.1) and discuss the relevance of
these results to the solar system (section 8.2).

\subsection{The $\upsilon$~Andromedae binary system}
Three planets orbit $\upsilon$~Andromedae \citep{b41}, two of which have their
apsides aligned \citep{b37,b38,b39}.  The most recent observations of this
system \citep{b40} yield the following orbital elements $a_b=0.059$\,AU,
$e_b=0.020$, $\omega_b=241^\circ$, $m_b=0.75\,m_{\rm J}$, $a_c=0.821$\,AU,
$e_c=0.185$, $\omega_c=214^\circ$, $m_c=2.25\,m_{\rm J}$, $a_d=2.57$\,AU,
$e_d=0.269$, $\omega_d=247^\circ$, and $m_d=2.57\,m_{\rm J}$ where the masses
are line-of-sight values, the stellar mass is $M=1.3M_\odot$.

Using the numerical integration of the full equations of motion (\ref{motion})
to model the eccentricity excitation with a constant-direction acceleration
applied to the planetary orbits with their current semi-major axes but with
initially circular co-planar orbits, we find that mutual planetary
perturbations are strong enough to damp the excitation if the acceleration is
smaller than $A_0\sim 10^{-11}{\rm km\,s}^{-2}$. The equivalent keplerian
boundary is at $a_{\rm kplr}\sim 500\,$AU. Below this value, the present
configuration can be recovered along with the apsidal alignment of the outer
two planets. However, the stellar companion to $\upsilon$~Andromedae
\citep{b36} puts an additional constraint on the excitation mechanism. At a
projected distance of 750\,AU, the presence of this $0.2M_\odot$ companion
leaves us two options: either the excitation by acceleration is ruled out
because the companion lies far outside the keplerian boundary of the weakest
possible acceleration that reproduces the planets' eccentricities, or that it
was initially inside the boundary and migrated by the sudden excitation that
we discussed in section 6. We point out that the projected distance of 750\,AU
does not translate necessarily into a semi-major axis as the companion's orbit
is likely to be eccentric.

\begin{figure}\epsscale{.7}
\plotone{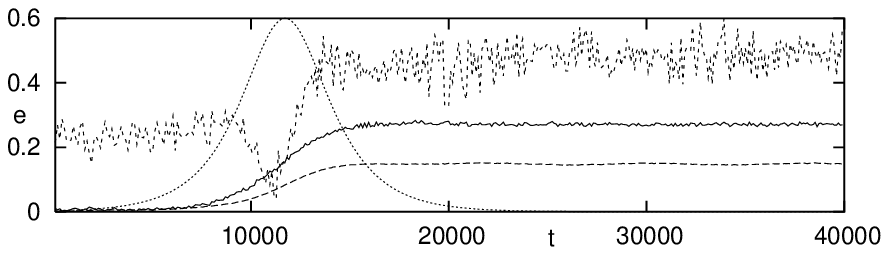}\\
\plotone{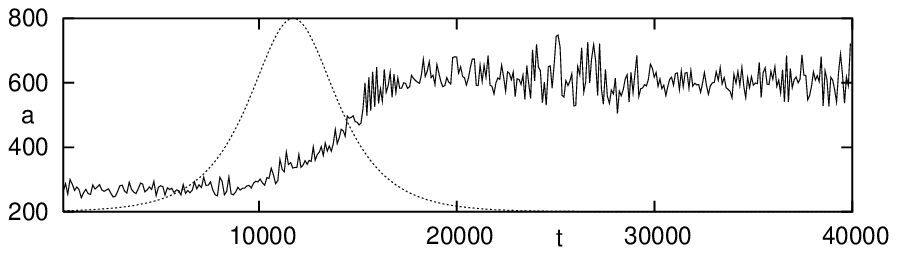}\\
\plotone{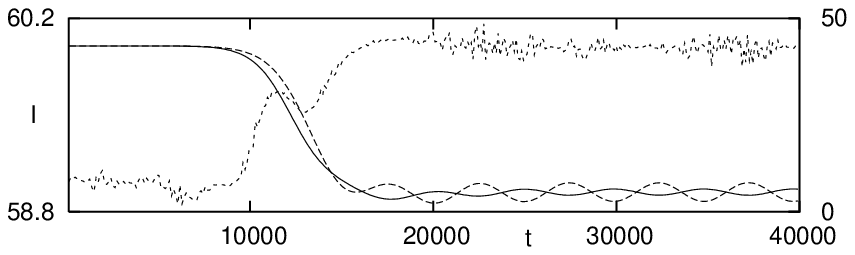}\\
\plotone{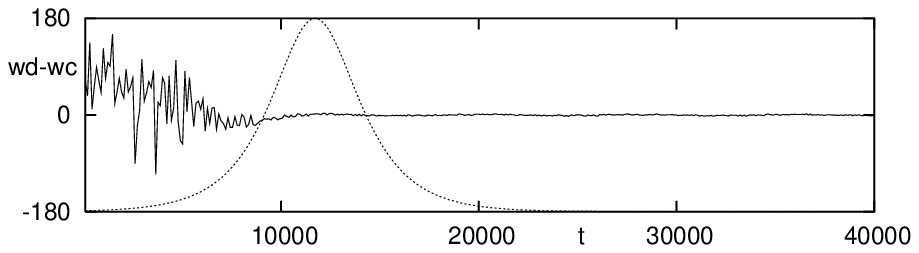}
\caption{Eccentricity excitation,  apsidal alignment and binary migration 
in the $\upsilon$~And system. The acceleration is of the form
$A_2$  (dotted) with $A_0=2.8\times
10^{-11}\,$km\,s$^{-2}$ , $\tau=2000$\,years, $t_0=12000$\,years corresponding
to $a_{\rm kplr}=300\,$AU and a residual velocity
$V=5.6$\,km\,s$^{-1}$. From the top down, 
The first panel shows the eccentricity excitation of
planets $d$ (solid) and  $c$ (dashed) and the eccentricity evolution of
$\upsilon$~And B (short-dashed) as well as the acceleration pulse
normalized to its maximum value (dotted).  The second panel shows the
radial migration of $\upsilon$~And B. The third panel shows the inclination 
of planets $d$ (solid, left scale) and  $c$ (dashed, left scale) 
and the inclination  evolution of $\upsilon$~And B (short-dashed, 
right scale). The last panel shows the relative apsidal libration of planets
$d$ and $c$.}
\label{fig-14}
\end{figure}

To test the second option, we choose an acceleration of the form $A_2$ with a
conservative keplerian boundary at $300$\, AU corresponding to $A_0\sim
3\times 10^{-11}$\,km\,s$^{-2}$, a duration $\tau=2000$\,years and an
equivalent residual velocity $V=5.6$\,km\,s$^{-1}$.  The planetary angular
momentum is inclined by $60^\circ$ with respect to the direction of
acceleration. The stellar companion's initial orbit has a semi-major axis
$a=298$\,AU, an eccentricity $e=0.3$ and an inclination of $10^\circ$ with
respect to the direction of acceleration. At this large semi-major axis, the
stellar companion's perturbation of the planets is negligible; in particular,
the eccentricity excitation by the Kozai mechanism \citep{b8} is not efficient
because the corresponding excitation time ($\sim 10^7$\,years) is much larger
the duration of acceleration and the eccentricity secular frequency of the
isolated two-planet system ($\sim 7000$\, years). We have chosen such an
eccentric and inclined stellar orbit to decouple the problem of eccentricity
excitation for the planetary and stellar companions.  We find that the
acceleration produces a configuration similar to the observed one with stellar
orbital elements: $e=0.5$, $a=600$\,AU, $I=50^\circ$ (Figure \ref{fig-14}). In
particular, we remark that the apsidal alignment that is generic in the basic
mechanism of section 3 for two bodies is maintained in the $\upsilon$
Andromedae system. We also note that only when the acceleration's strength is
near maximum and the keplerian boundary nears 300\,AU, does the orbit acquire
a larger eccentricity. The numerical integration also confirms the absence of
the Kozai eccentricity perturbations in the orbits' evolution as suggested by
the timescale analysis.

The acceleration's magnitude of $A_0\sim 3\times 10^{-11}$\,km\,s$^{-2}$
corresponds to a jet-driven mass loss rate $\dot M\sim
10^{-6}M_\odot$\,year$^{-1}$. This value is within the two orders of magnitude
uncertainty in mass loss estimation (section 2). However, the inferred rate is
a maximal value in time (e.g. Figure \ref{fig-14}) and cannot be compared
directly to the observed steady state values which are naturally smaller. An
additional constraint on the excitation by jet acceleration is that the planet
must not lie far inside the resonance location.  In terms of the parameters we
introduced, the resonant semi-major axis of the precessing jet is given as:
\begin{equation}
a_{\rm res}\simeq 4 \ \ \left(\frac{M_\odot}{M+m}\right)\,\left(\frac{a_{\rm
 kplr}}{10^2\, {\rm AU}}\right)^4\left(\frac{10^4\,\mbox{years}}{T_{\rm
 prec}}\right)^2\ \ \mbox{AU}.
\end{equation}
where $T_{\rm prec}=2\pi/\Omega_A$ is the jet's precession period. This
estimate shows that the most efficient accelerating jets precess with periods
of $10^4$ years or larger.  If the precession angle is large as it is the case
here, then the large eccentricity region is extended inside the resonant
semi-major axis by a sizable factor (e.g. Figure \ref{fig-5} with
$\alpha=30^\circ$).  At $a=a_{\rm res}/100$, maximum amplitude is 0.17 for
$\alpha=60^\circ$.  This simple example allows us to conclude that
stellar-disk mass loss is a possible process to provide acceleration. Further
modeling of jet acceleration with a more realistic amplitude time variation is
needed to constrain the excitation mechanism.

\subsection{The solar system}
The solar system was used in our model to motivate the possible options of the
direction of acceleration using the small eccentricities of Jupiter and
Saturn. The detailed analysis of the excitation by acceleration pointed out
other ways to limit the eccentricity growth: to the maximum eccentricity set
by the inclination of the acceleration with respect to the orbital plane, we
add the amplitude reduction associated with a precessing acceleration and that
associated with a weak acceleration ($a_{\rm kplr}>2600\,$AU with
$I_0=30^\circ$ constant) that competes with mutual planetary perturbations.
These options show that it is more difficult to identify the evolutionary
track of low eccentricity systems. Modeling the migration of the solar
system's planets and its effects on minor bodies under a perturbing
acceleration may offer a way to discriminate between the available options.
The current state of the giant planets' orbits is believed to have evolved
from a much more compact system. Studies of the origin of the eccentricity and
inclination distributions of Kuiper belt objects suggest that the four giant
planets were confined between 5.5\,AU and 13.5\,AU at the time where the
Kuiper belt was made up of a low eccentricity and inclination planetesimal
disk \citep{b14,b33}. For such compact configurations, mean motion resonance
crossing may produce eccentricities comparable to the current values for
Jupiter and Saturn \citep{b44}. Regardless of the specifics of such mechanisms
and if we admit that accelerating processes of the type studied here excited
companion eccentricities for solar-type stars in the solar neighborhood, the
same processes must have applied to the solar system and shaped it to a
certain extent. In the context of acceleration by stellar jets and star-disk
winds, the inclination of Jupiter's angular momentum vector by $6^\circ$ with
respect to the Sun's rotation axis is an indication that Jupiter's orbital
plane has had to evolve from an equatorial accretion disk or that the planet
had formed in a warped disk. Both possibilities are favorable to the
excitation by acceleration model.

An interesting application of our model that does not require modifying the
standard picture of how the solar system and its minor body populations have
evolved is to choose the direction of acceleration nearly perpendicular to the
planetary orbital plane. The acceleration, however, may not be particularly
weak thereby allowing smaller values of $a_{\rm kplr}$.  The main consequence
of such an acceleration is to excite the eccentricities of bodies that lie
near the keplerian boundary, eject those that lie far outside it, thus
truncating the planetesimal disk, and outward transport those in its vicinity
through the migration that we discussed in section 6 and illustrated in
Figures (\ref{fig-9}) and (\ref{fig-10}).  Migration could enhance the
delivery of minor bodies to the Oort Cloud and explain the transport of Kuiper
Belt outliers 2000 CR105 and Sedna (90377) that, with perihelia larger then
Neptune's semi-major axis, still elude dynamical explanation \citep{b58,b45}.

\section{Concluding remarks}
In this paper, we have worked towards building a theory for the origin of the
eccentricities of extrasolar planets. Proceeding with minimal assumptions, we
showed that the planetary eccentricities can be caused by relative
accelerations that depend weakly on the local dynamics. We have thus reduced
the problem of the origin of eccentricities to the identification of the
physical processes that cause such accelerations. Possible processes are
stellar jets and star-disk winds that accelerate the host star with respect to
the companion and excite the random component of its galactic velocity.  The
origin of extrasolar eccentricities can therefore be related to the the random
motion of disk stars in the Galaxy.

The model has further applications to the dynamics of extrasolar planets and
the solar system. If the duration of acceleration is larger than the
excitation time at some semi-major axis, planets exterior to this radius will
achieve maximum eccentricity and risk ejections by close encounters (section
3). This offers a possible explanation for single planet systems with large
eccentricities. Rogue planets can also be produced if planets are pushed out
to the keplerian boundary by disk-planet interaction (section 6). For the
solar system, besides the possible eccentricity excitation, acceleration leads
to the transport of minor bodies to the outer Kuiper Belt and the Oort Cloud.

The excitation by acceleration model can be applied in other contexts beside
the stellar and planetary and companions.  The star-disk mass loss mechanisms,
that can be responsible for the eccentricity of extrasolar planets, impart an
equal acceleration to the protoplanetary disk. A precessing acceleration
derives from the perturbing potential $R={\bf A}\cdot {\bf x} = r A_r \cos
(\theta-\theta_A) + z A_z$ where $\theta_A$, $A_r$, $A_z$ are the
acceleration's components in a cylindrical coordinate system referred to the
disk. This potential could be able to excite the $m=1$ slow modes of the disk
\citep{b28}. Such modes are interesting because of their large wavelength that
can be comparable to the size of the system. For instance, the disk could
develop rigid precession if the speed of sound waves is larger than the
precession rate similarly to the case of tilted disks perturbed by a binary
component \citep{b42}. The feedback of a jet-generated precessing acceleration
on the disk may have important consequences for sustaining the acceleration.

The excitation mechanism also applies to galactic dynamics.  Large-scale wind
phenomena \citep{b2,b21} accelerate the Galaxy and alter the galactic rotation
curve by decreasing the circular velocity. The analysis of this paper can be
modified to ascertain the efficiency of galactic winds in the excitation of
the stellar random motion by substituting the galactic potential for the
stellar potential in equation (\ref{motion}). Further applications of galactic
winds include the stability of large scale structures in disk galaxies, the
onset of $m=1$ elliptic distortions, and the effect of the change in the Sun's
Hill sphere in the Galaxy on the dynamics of the Oort Could.

\section*{Acknowledgments}
I thank an anonymous referee for the careful review of the paper and Rodney
Gomes, Fran\c cois Mignard and Kleomenis Tsiganis for discussions. This work
was supported by the Programme National de Plan\'etologie.

\appendix
\section[]{Averaged potential of a constant acceleration}
In the following, we derive the averaged potential of a constant acceleration
${\bf A}$. The acceleration (or equivalently its force) derives from the
potential $R={\bf A}\cdot {\bf x}$ where ${\bf x}$ is the position vector. The
averaged potential is $\left<R\right>={\bf A}\cdot \left<{\bf x}\right>$ and
therefore only the average of the position vector ${\bf x}$ needs to be
calculated. The companion's unperturbed orbit is keplerian and in an
orthonormal basis can be written as:
\begin{eqnarray}
x&=& r \left(\cos\Omega \cos(\omega+f)-\sin\Omega\sin(\omega+f)\cos I\right), 
\label{x}\\
y&=& r \left(\sin\Omega \cos(\omega+f)+\cos\Omega\sin(\omega+f)\cos I\right),
\label{y}\\
z&=& r \sin(\omega+f)\sin I, \label{z}
\end{eqnarray}  
where the radius is $r=a(1-e^2)/(1+e\cos f)$ and the orbital elements $a,\ f,\
e,\ \omega, I$, and $\Omega$ are respectively, the semi-major axis, the true
anomaly, the eccentricity, the argument of pericenter, the inclination and the
longitude of the ascending node. The time average can be replaced with a true
anomaly average by using the conservation of angular momentum ${\rm d} t/P=
r^2 {\rm d} f/2\pi a^2\sqrt{1-e^2}$, where $P$ is the period. Applying this
integration to the position vector gives:
\begin{equation}
\left<{\bf x}\right>=-\frac{3}{2}\, a \, e\, \frac{\bf x}{r}(f=0). \label{xmean}
\end{equation}
The direction of ${\bf x}$ at $f=0$ is that of the eccentricity vector (or
Runge-Lenz vector), ${\bf e}={\bf v}\times {\bf h}/G(M+m)-{\bf x}/r$. The
averaged potential is therefore:
\begin{equation}
\left<R\right>=-\frac{3}{2}\, a \, {\bf A}\cdot{\bf e}. \label{secpotA}
\end{equation}
Inspection of the position vector equations shows that the simplest expression
corresponds to choosing the $z$-axis as the direction of acceleration. This
leads to $\left<R\right>=-3aeA\sin(\varpi-\Omega)\sin I/2$ where
$\varpi=\omega+\Omega$ is the longitude of the pericenter, the conjugate
variable of the eccentricity, $e$.

%%%   bib counter at b58

%%%%%%%%%%%
%%%  FIG  1

%%%%%%%%%%%
%%%  FIG  2

%%%%%%%%%%%
%%%  FIG  3  

%%%%%%%%%%
%%% FIG  4

%%%%%%%%%%
%%% FIG  5

%%%%%%%%%%
%%% FIG  6

%%%%%%%%%%
%%% FIG  7

%%%%%%%%%%%
%%%  FIG 8

%%%%%%%%%%%%
%%%  FIG 9

%%%%%%%%%%%%%
%%%  FIG 10

%%%%%%%%%%%
%%%  FIG 11

%%%%%%%%%%%
%%%  FIG 12

%%%%%%%%%%%
%%%  FIG 13

%%%%%%%%%%%

%%%%%%%%%%%
%%%  FIG 14

\end{document}